\documentclass[preprint,12pt,sort&compress]{elsarticle}

\usepackage{bm}
\usepackage{mathrsfs}
\usepackage{amsmath}
\usepackage{amsfonts}
\usepackage{amsthm}
\usepackage{color}
\usepackage{graphicx}
\usepackage{enumitem}
\usepackage{tikz}
\usepackage{xcolor}
\usetikzlibrary{shapes.geometric, arrows}
\tikzstyle{startstop} = [rectangle, rounded corners, text centered, draw=black]
\tikzstyle{process} = [rectangle, text centered, draw=black]
\tikzstyle{decision} = [diamond, text centered, draw=black]
\tikzstyle{arrow} = [thick,->,>=stealth]

\usepackage{caption}
\usepackage{subcaption}
\usepackage{float}
\usepackage{url}
\usepackage{multicol}
\usepackage[top=1in,bottom=1in,left=1in,right=1in]{geometry}
\usepackage[pdfborder={0 0 0},colorlinks,allcolors=blue]{hyperref}
\usepackage{setspace}
\usepackage{arydshln}
\usepackage{booktabs}
\usepackage{verbatim}

\theoremstyle{definition}%plain/definition/remark
\graphicspath{{figures/}}

\allowdisplaybreaks

\begin{document}
%\numberwithin{equation}{section}

\begin{frontmatter}



\title{Machine learning for metal additive manufacturing: Predicting temperature and melt pool fluid dynamics using physics-informed neural networks}

\author[label1]{Qiming Zhu}
\author[label2]{Zeliang Liu\corref{cor1}}
\author[label1]{Jinhui Yan\corref{cor1}}
%\ead{yjh@illinois.edu}

\address[label1]{Department of Civil and Environmental Engineering,\\ University of Illinois at Urbana-Champaign, Champaign, IL, USA}
\address[label2]{Livermore Software Technology,\\ an ANSYS company, Livermore, CA, USA}

\cortext[cor1]{Corresponding author}

\begin{abstract}

The recent explosion of machine learning (ML) and artificial intelligence (AI) shows great potential in the breakthrough of metal additive manufacturing (AM) process modeling, which is an indispensable step to derive the process-structure-property relationship. However, the success of conventional machine learning tools in data science is primarily attributed to the unprecedented large amount of labeled data-sets (big data), which can be either obtained by experiments or first-principle simulations. Unfortunately, these labeled data-sets are expensive to obtain in AM due to the high expense of the AM experiments and prohibitive computational cost of high-fidelity simulations, hindering the direct applications of big-data based ML tools to metal AM problems.

To fully exploit the power of machine learning for metal AM while alleviating the dependence on ``big data", we put forth a physics-informed neural network (PINN) framework that fuses both data and first physical principles, including conservation laws of momentum, mass, and energy, into the neural network to inform the learning processes. To the best knowledge of the authors, this is the first application of physics-informed deep learning to three dimensional AM processes modeling. Besides, we propose a hard-type approach for Dirichlet boundary conditions (BCs) based on a Heaviside function, which can not only exactly enforce the BCs but also accelerate the learning process. The PINN framework is applied to two representative metal manufacturing problems, including the 2018 NIST AM-Benchmark test series. We carefully assess the performance of the PINN model by comparing the predictions with available experimental data and high-fidelity simulation results, , using finite element based variational multi-scale formulation. The investigations show that the PINN, owed to the additional physical knowledge, can accurately predict the temperature and melt pool dynamics during metal AM processes with only a moderate amount of labeled data-sets. The foray of PINN to metal AM shows the great potential of physics-informed deep learning for broader applications to advanced manufacturing. All the data-sets and the PINN code will be made open-sourced in https://yan.cee.illinois.edu/ once the paper is published.
\end{abstract}

%\begin{keyword}
%Machine learning, PINN, AM-Benchmark, metal AM
%\end{keyword}

\end{frontmatter}

%\addtocontents{toc}{\protect\sloppy}https://www.overleaf.com/project/5ed849300d65ce00013f61cd
%\tableofcontents
%\newpage
%\linenumbers
%\onehalfspacing

% List of figures

%%-------------------------------------------------------------------------------------------------
\section{Introduction}
\label{sec:introduction}
%%-------------------------------------------------------------------------------------------------
It has been widely believed that metal additive manufacturing (AM)  can revolutionize mechanical, aerospace, and biomedical industries owing to its superior capability to print metals with complex geometries directly from digital models without the design constraints of traditional manufacturing routes. The market of metal AM has been growing significantly for the past decade. However, compared with the total manufacturing market, the industrial adoption of metal AM has not reached its expected potential due to a lack of reliable process-structure-property relationships.  For the past several years, predictive computational models, in conjunction with in-situ and ex-situ measurements and monitoring~\cite{exp0,exp1,exp2}, have been playing an indispensable role in enhancing the understanding of the process-structure-property relationship in metal AM. Federal agencies have also conducted several benchmark experiments, such as NIST AM-bench~\cite{NISTAM} and AFRL AM modeling challenge series~\cite{AFAM}, to facilitate the development of metal AM modeling tools.

Among various computational models at different scales and fidelity, thermal-fluid process simulation is not only an essential tool to understand the metal AM physics but also acts as a spearhead to derive the process-structure-property relationship. The metal AM process is intrinsically a multi-scale and multi-physics problem, involving rapid, complex, and coupled mass/flow/heat exchanges between gas, liquid, and solid phases, with large density ratios and complicated interfacial phenomenon. Current numerical simulation tools often employ mathematical models that couple Navier-Stokes equations and a heat transfer equation to capture the evolution of temperature and melt pool dynamics during manufacturing processes. For decades, the manufacturing community has been adopting computational methods that directly solve these mathematical models or their weak forms, based on spatial discretization (e.g., finite difference, finite volume, finite element, and mesh-free methods) and time-stepping.

The predictive capacities of these approaches have been significantly enhanced, thanks to the researchers' persistent efforts in numerical method development from the manufacturing and computational mechanics/mathematics communities. \textcolor{black}{For example, Lawrence Livermore National Lab developed a thermal-fluid solver using the Arbitrary-Lagrangian Eulerian technique, which can simulate laser powder bed fusion (LPBF) processes at powder-scale~\cite{ALE3D0,ALE3D1, ALE3D2, ALE3D3}; Knapp et al.~\cite{DED_Debroy} and Mukherjee et al.~\cite{LPBF1, LPBF2} developed a coupled thermal-fluid model to simulate directed energy deposition (DED) and laser powder bed fusion (LPBF) processes. Lin et al.~\cite{lin2019numerical,Lin2020} developed a control-volume finite element approach to simulate directed energy deposition process. Lattice Boltzmann method has been used to model the metal powder melting and re-solidification in~\cite{attar2011lattice, korner2011mesoscopic,korner2013fundamental}; Zohdi group employed a discrete particle method to describe the selective laser sintering process~\cite{zohdi2014additive,zohdi2014direct,ganeriwala2014multiphysics}; Yan et al. developed a volume-of-fluid (VoF) based thermal-fluid solver to model multi-layer and multi-track LPBF process~\cite{yan2017multi,yan2018meso, wentao0,wentao1, wentao2}; Panwisawas et al. also employed a VoF method by using OpenFOAM to analyze the inter-layer and inter-track void formation~\cite{panwisawas2017mesoscale}; Li et al. developed a thermal-fluid model by combining level set method and Lagrangian particle tracking to investigate powder-gas interaction in LPBF processes~\cite{wenda0}. CFD-ACE+, a code developed by ESI group, has been used to analyze the defects such as porosity, balling, and denudation in metal AM~\cite{megahed2016powder,mindt2018modeling}; The last author of this paper~\cite{yan2018fully} developed a gas-liquid-solid thermal flow model based on the level set method and residual-based variational multi-scale method to simulate laser spot melt pool flows. Li. et al. used a mesh-free model based on material point method for selective laser beam melting processes~\cite{MeshfreeAM}. Gan et al. developed a finite element method (FEM) based thermal-fluid model and applied it to the NIST AM-Bench problems~\cite{gan2019benchmark}.} 

 %Hojjatzadeh et al. used a multi-phase and multi-physics model to investigate the pore formation and elimination~\cite{hojjatzadeh2019pore};
The core of these conventional approaches can be summarized as the process of replacing PDEs with a set of algebra equations that can be handled by digital computers. Despite the continued success and evolution, these methods require sophisticated mathematical treatments for spatiotemporal discretizations, coupling strategies, boundary conditions, and linear solvers to ensure stability, robustness, and efficiency. The application of these approaches to real additive manufacturing problems is prohibitively expensive and intricate. The high-fidelity simulations are typically executed in a parallel environment and consume massive high-performance computing (HPC) hours. Also, the performance of these approaches is often problem-dependent, necessitating numerical practitioners to have a deep understanding of not only the manufacturing problems but also the underlying mathematical techniques.

Machine learning (ML) and artificial intelligence (AI) have the potential to accelerate breakthroughs in thermal-fluid modeling for metal AM processes by harnessing data from sensors, experiments, and high-fidelity simulations. In general, ML focuses on algorithmic modeling of data and making predictions of labels based on observations, with emphasis on making accurate predictions for classification and regression tasks. Modern deep learning approaches have demonstrated tremendous successes in domains ranging from sentiment analysis to chemical predictions to material design~\cite{liu2019transfer, liu2019deep, liu2019exploring,liu2018data,liu2018microstructural,liu2016self}. The first reason for the major success of modern ML techniques, especially deep learning, is the availability of vast amounts of data (big-data). The second reason is that many technical burdens have been mitigated by advances in both hardware and software, including high-performance computers, graphics processing units (GPUs), fast large-scale optimization schemes, new optimality guarantees, and many user-friendly open-sourced packages, such as Tensorflow~\cite{abadi2016tensorflow}, PyTorch~\cite{paszke2019pytorch}, Theano~\cite{bastien2012theano}, and Caffe~\cite{jia2014caffe}.

However, using deep learning for AM process modeling is still challenging. The primary challenge arises from the lack of large labeled data-sets since either experimental measurements or high-fidelity simulated data of AM processes are expensive to attain, rendering the big data-based ML/AI algorithms infeasible. The good news in scientific problems, however, is that there is highly condensed knowledge and expertise available in fundamental conservation, evolution, or constitutive principles, which are often expressed as a set of partial differential equations (PDEs). One can incorporate this type of knowledge into ML/AI models to enhance their predictive capability in sparse data regions. Nowadays, these approaches are coined as scientific machine learning (SciML) in the computational mathematics/mechanics communities. In particular, a widely used approach in SciML is to train a conventional deep learning (DL) model such as Gaussian process regression (GPR)~\cite{Gauss1,raissi2017machine} or deep neural network (DNN)~\cite{FirstPINN, PINN,sun2020surrogate,zissis2015cloud,raissi2017physics} with physical principle constraints. Existing research has demonstrated SciML’s capability in sparse-data scenarios for various application areas, such as environmental study~\cite{PINN5, PINN6}, material science~\cite{lu2020extraction, PINN4}, and cardiovascular modeling~\cite{PINN,PINN1,PINN3}.

\textcolor{black}{Although most of the SciML applications are restricted to single physics systems, we envision the general concept can be extended to tackle the multi-physics problems in metal AM. Thus, this paper put forth a SciML framework for metal additive manufacturing processes to predict the temperature field and melt pool fluid dynamics via a physics-informed neural network (PINN). We aim to fully take advantage of the prediction capabilities of deep neural networks while significantly reducing the amount of costly training labeled data. To this end, the physical conservation laws of momentum, mass, and energy are fused into a fully connected neural network by penalizing the loss function with the residuals of the Navier-Stokes equations and enthalpy conservation equation on a set of collocation points. Owed to this additional knowledge, the learning process only requires a small amount of labeled data-set. Besides, to impose the necessary Dirichlet boundary condition (BC), we borrow the idea from the interface-capturing approach widely used in multi-phase fluid mechanics, in which a small portion of the neural network is solely used to enforce the Dirichlet BC by a Heaviside function. This ``hard" approach can not only precisely satisfy the Dirichlet BC but also speed up the learning process, compared with the conventional ``soft" approach that uses additional constraint in the loss function to enforce the BC. Once the model is trained, the quantities of interest, such as temperature, velocity, pressure, and melt pool dimensions, can be predicted accurately.
}

The paper is structured as follows. Section~\ref{sec:method} represents the physics-informed neural network framework, in which the PDEs of physical principles, design of loss function, enforcement of Dirichlet BC, and training procedures are introduced in an articulated way. Section~\ref{sec: Applications} demonstrates the applications of the PINN framework to two representative manufacturing problems. The first application is using the PINN framework to solve a classic solidification problem from the textbook by Dantzig and Rappaz~\cite{dantzig2016solidification}. For this problem, the PINN is informed by the energy conservation law and trained without labeled data-set. We compare the performance of ``hard" BC and ``soft" BC on this problem in terms of both accuracy and learning efficiency. The predictive capability of the PINN is assessed by comparing it with the standard finite element method (FEM) with resolution refinement studies. The second application is utilizing the PINN framework to predict the temperature field and melt pool fluid dynamics for the 2018 NIST AM-Bench test series. We utilize a validated finite element based variational multi-scale formulation (VMS)~\cite{yan2018fully} to generate the synthetic training data-sets. The investigations show that the PINN, informed by conservation laws of momentum, mass, and energy, can accurately and efficiently predict the melt pool dimension, fluid field, and cooling rate for the three selective laser beam melting tests done by NIST with a small amount of training data. We summarize the contributions and limitations of the paper and outline future work in Section~\ref{sec:concl}. The methods and setup details of high-fidelity FEM based thermal fluid simulations that are used for generating training data and validation are presented in the \textcolor{blue}{Appendix}.

\section{Methods}
\label{sec:method}
\subsection{Governing partial differential equations}

This section presents the governing equations of the thermal-fluid flows in metal AM processes. The theory of the equations builds upon the tacit assumptions that \textcolor{black}{the solid phase is a highly viscous fluid with the same constant density as the liquid phase,} and the loss of metal material due to vaporization~\cite{vapor1,vapor2} and the effects on heat loss, composition change and fluid motion are 
negligible. A flat top surface is adopted based on the fact that the melt pool deformation is small compared with the melt pool dimensions in the problems considered in the paper. With the above assumptions, the thermal-fluid model based on conversation laws of momentum, mass, and energy is defined as the following coupled PDEs
\begin{align}
&              \rho ( {\bm{u}}_{,t} +  \bm{u}\cdot \nabla \textbf{u}-\bm{g}) + \nabla p - 2\mu \Delta \bm{u}  = \textbf{0}  \label{eq:U_momentum_equ}\\
&    \nabla\cdot \bm{u} = 0  \label{eq:U_continuity_equ} \\
&       (\rho c_p T)_{,t} + \bm{u}\cdot \nabla (\rho c_pT) + (\rho L f_L)_{,t}  +  \bm{u}\cdot \nabla (\rho L f_L) - \kappa \nabla^2 T - Q_T  = 0  \label{eq:Temp_equ}
\end{align}

Here Eq.~\ref{eq:U_momentum_equ} and Eq.~\ref{eq:U_continuity_equ} are the Navier-Stokes equations of incompressible flows, where $\bm{u}$ is the velocity field, $p$ is the pressure field, $\textbf{g}$ is the gravitational acceleration vector, $\nabla$ is the gradient operator, $\Delta$ is the Laplace operator, $\rho$ and $\mu$ are the density and dynamic viscosity, respectively.

Eq.~\ref{eq:Temp_equ} is the conversation equation of energy, where $T$ is temperature field, $c_p$ is the specific heat capacity, $L$ is the latent heat of fusion, $\kappa$ is the thermal conductivity, $Q_T$ is an energy source. 

To have well-posed systems,  Eqs.\ref{eq:U_momentum_equ}-\ref{eq:Temp_equ} are subjected to the following Dirichlet and Neumann boundary conditions

\begin{align}
&    \bm{u} = \bm{u}_{bc}  \label{U_bc}\\
&     {p} = {p}_{bc}    \label{P_BC} \\
&      {T} = {T}_{bc}  \label{T_BC} \\
&     -p\bm{n}+2\mu \nabla^S \bm{u} \cdot \bm{n} = \bm{\tau}  \label{Traction_BC}\\
&  \kappa \nabla T \cdot \bm{n} = q \label{Flux_BC}
\end{align}

where $\bm{u}_{bc}$, ${p}_{bc}$ and ${T}_{bc}$ are the prescribed velocity, pressure, and temperature on Dirichlet boundaries, respectively. $\bm{\tau}$ and $q$ are the prescribed traction and heat flux on the Neumann boundaries, respectively. $\nabla ^S$ is a symmetric gradient operator and $\bm{n}$ is the unit normal vector on the boundary.  

In the model, the solid and liquid phases are distinguished by a liquid fraction $f_L$, which takes 1 in the liquid phase, 0 in the solid phase, and a linear profile in the mushy zone~\cite{voller1991eral}. $f_L$ is defined as 
\begin{align}
    \label{fl_formula}
    f_L =\left\{
           \begin{array}{ll}
              0 \qquad &if \ T < T_s \\
              \frac{T-T_s}{T_l-T_s} \qquad &if\ T_s \leq T \leq T_l\\
              1 \qquad &if \ T > T_l
            \end{array}
            \right.
\end{align}
where $T_s$ and $T_l$ are the solidus and liquidus temperature, respectively. 

 With the assistance of $f_L$, the material properties in the thermal-fluid model are evaluated by the following interpolation
\begin{align}
 \label{var_interpolate}
 \psi = f_L \psi_L + (1-f_L)\psi_S
\end{align}
where $\psi$ denotes the specific material property in the model (e.g., density, dynamic viscosity, specific heat capacity, heat conductivity), and $\psi_L$ and $\psi_S$ are the corresponding property in the liquid and solid phase, respectively. 
%To inhibit the fluid motion in the solid phase, there are two options for numerical simulation. The first option is to use a very large viscosity for the solid phase. If the magnitude of the viscosity of solid is too large, the conditioning of the linear system would be bad. The second option is to add a Darcy force term based on Carman-Kozeny model \cite{yan2018fully,voller1987enthalpy} in the momentum equation. The first approach is used in the data generation of FEM simulation.
\subsection{Physical informed neural network (PINN) for Thermal-fluid flows}
%\textcolor{black}{\subsubsection{Neural network}}
%\textcolor{black}{The idea of the neural network is from a biological neural network in brains. There are three popular types of neural network: fully connected neural network (FC-NN) \cite{schwing2015fully}, convolutional neural network (CNN) \cite{lawrence1997face}, recurrent neural network (RNN) \cite{mikolov2010recurrent}.}

\begin{figure}[!htbp]
\centering
\includegraphics[scale=0.18]{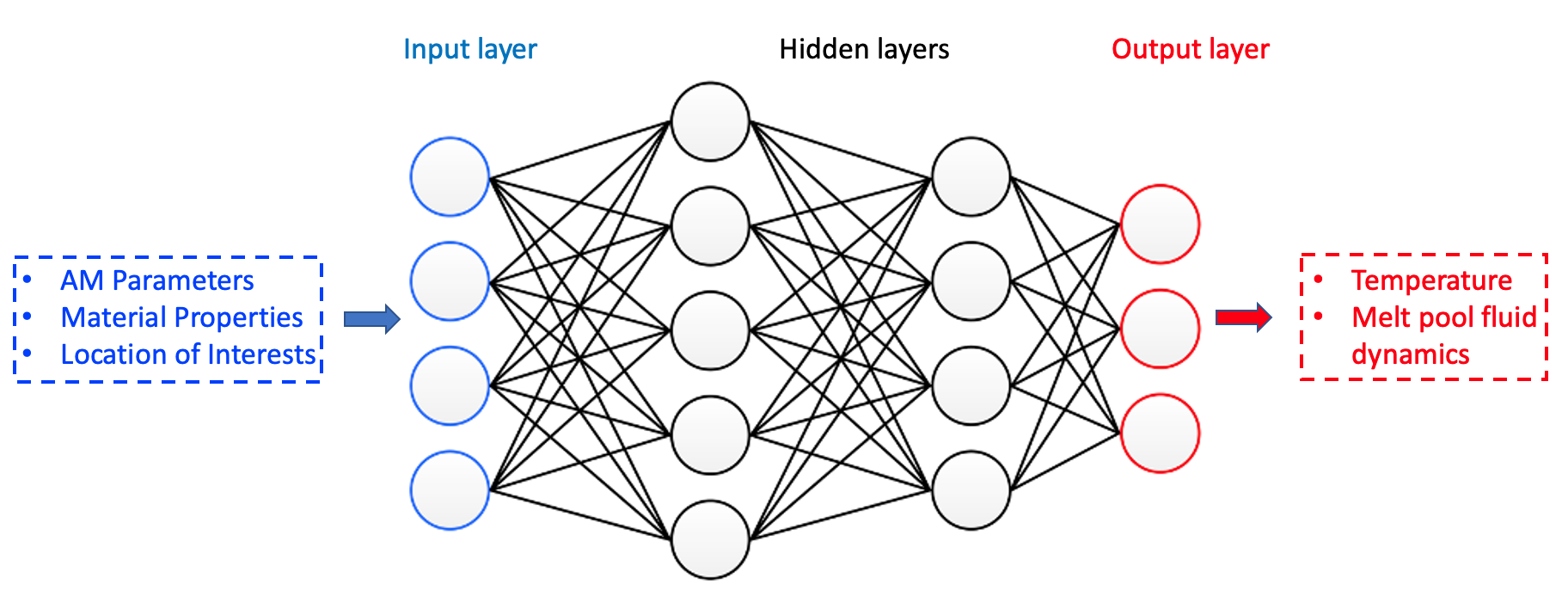}
\caption{A fully connected deep neural network for metal AM.}
\label{fig:NN_example}
\end{figure}

Neural network is a computing architecture that is vaguely inspired by the biological neural networks that constitute animal brains~\cite{liu2017survey}. Some typical neural network architectures are fully connected neural network (FC-NN) \cite{schwing2015fully}, convolutional neural network (CNN) \cite{lawrence1997face}, and recurrent neural network (RNN) \cite{mikolov2010recurrent}, which have been successfully used in a variety of machine learning applications, such as system identification and control, signal classification, pattern recognition, 3D reconstruction, sequence recognition, social network filtering, data mining, and medical diagnosis~\cite{zissis2015cloud, sengupta2016lung,bishop2006pattern,choy20163d,han2011data}. In this paper, the PINN of the thermal-fluid model makes use of a fully connected deep neural network (FCNN)~\cite{{schwing2015fully}}, where the neurons of adjacent layers are fully connected. Figure~\ref{fig:NN_example} shows the schematic picture of the fully connected neural network used in this paper, which consists of an input layer, hidden layers, and an output layer. A neural network with more than one hidden layer is conventionally called a deep neural network, whose function approximation capability increases with the number of hidden layers and neurons~\cite{liang2016deep}. A deep neural network maps the input $\bm{z_0}$ to the output $\bm{z_L}$ from the input layer to the output layer. In the hidden layers, each layer receives outputs from the previous layer and feeds forward inputs to the next layer. 
The relation of between the input and output of the $l^{th}$ layer ($l= 1,...,L-1$) is defined as 
\begin{align}
 \label{NN_prop}
 \bm{z}_l = \sigma_l(\textbf{w}^T_l {\bm{z}}_{l-1} + \textbf{b}_l)
\end{align}

where $\textbf{w}_l$ and $\textbf{b}_l$ are the weight matrix and bias vector of this layer, respectively. $\sigma_l$ is the activation function that can introduce the non-linearity to the system~\cite{sibi2013analysis}. Widely used activation functions in deep learning are tanh function, Rectified linear unit (Relu), sigmoid function~\cite{maas2013rectifier}. In this paper, we employ a swish activation function~\cite{eger2019time}, which is a smoothed version of Relu function, defined as
\begin{align}
\label{NN_swish}
\sigma (x) = swish(x) = x\ sigmoid(x)=x/(1-e^{-x})
\end{align}

%In this paper, we construct a fully connected neural network to represent the solutions of metal AM processes (the solutions of PDEs in Eqs.~\ref{eq:bulk} subjected to the boundary conditions in Eqs.~\ref{eq:BC}). 

The goal of the neural network is to learn the following mapping for a given set of manufacturing parameters (e.g,. alloy properties, laser power, and scanning speed): 
\begin{align}
\label{eq: mapping}
[t,\bm{x}] \xrightarrow[] {\bm{W}, \bm{b}} [\bm{u}_{NN}, p_{NN} , T_{NN}] 
\end{align}
where the input $[t,\bm{x}]$ are the collocation points (in both space and time) of interest. The output $\bm{u}_{NN}, p_{NN}, T_{NN}$ are the velocity, pressure, and temperature fields we want to predict. To enable such a mapping, the \textcolor{black}{hidden parameters} of the neural network, $\bm{W}$ and $\bm{b}$, need to be identified by optimizing a meticulously designed loss function, which will be given in the next section. \textcolor{black}{Once $\bm{W}$, and $\bm{b}$ are determined, the output prediction can be easily achieved by a feed-forward evaluation, which is very efficient since only a few matrix multiplications are needed in Eq.~\ref{NN_prop}.}
%$\bm{z}_L$, such as velocity, pressure, and temperature in the current application, can be efficiently predicted from given input vector $\bm{z}_0$, such as collocation points (both space and time) of interest and manufacturing parameters in the current application, based on the Eq.~\ref{NN_prop}. The cost of this feed-forward evaluation is very low compared with transitional FEM methods since only a few matrix multiplications is needed in Eq.~\ref{NN_prop}.}
%For a problem with simple geometry, function $H(t,\bm{x})$ could be obtained by writing the analytic form. For a problem with complex geometry, function $H(t,\bm{x})$ could be obtained from solving Eikonal equation or training another small NN \cite{sun2020surrogate}.
\textcolor{black}{\subsubsection{Loss function design}}
The loss function, $L(\bm{W},\bm{b})$, in the PINN for thermal-fluid flows consists of two components: $L_{data}(\bm{W},\bm{b})$ and $L_{pde}(\bm{W},\bm{b})$, which represents the constraint of matching existing labeled data and the constraints of satisfying fundamental physical principles. Their definitions are given as follows.

Let $\hat{\bm{u}}$, $\hat{p}$, and $\hat{T}$ denote the available labeled data for velocity, pressure, and temperature, respectively. These labeled data can be either obtained by experiments or validated high-fidelity simulations. As a mean squared deviation (MSD) of the discrepancy between the prediction and labeled data, the component in the loss function from the data constraint, $L_{data}$, is defined as
\begin{align}
 \label{eq: Loss_data}
 L_{data}(\bm{W},\bm{b})= \frac{1}{N_{u}} \sum_{i=1}^{N_u} \left[  {\bm{u}_{NN}} (\bm{x}_i, t_i, \bm{W}, \bm{b}) -\hat{\bm{u}} (\bm{x}_i, t_i)  \right]^2  \\\nonumber 
 +\frac{1}{N_p} \sum_{i=1}^{N_p} \left[   {p}_{NN}(\bm{x}_i, t_i, \bm{W}, \bm{b}) - \hat{p} (\bm{x}_i, t_i) \right]^2 \\\nonumber 
 + \frac{1}{N_T} \sum_{i=1}^{N_T}\left[  {T}_{NN} (\bm{x}_i, t_i, \bm{W}, \bm{b}) - \hat{T} (\bm{x}_i, t_i)  \right]^2
\end{align}
where $N_{u}$, $N_{p}$, and $N_{T}$ are the number of labeled velocity, pressure, and temperature data points, respectively. Conventional off-the-shelf machine learning tools purely minimize this loss function to identify the hidden parameters. The success of this approach requires a massive amount of data-sets. However, considering the cost of experimental measurements and high-fidelity simulations, these labeled velocity, pressure, and temperature data points are expensive (sometimes impossible) to obtain. This limitation hinders the direct application of big-data based machine learning tools to metal AM process prediction.

To alleviate the dependence on big-data, we substitute extra expertise in fundamental physical principles into the loss function. These physical principles, often expressed as a set of PDEs with appropriate initial and boundary conditions, are highly condensed knowledge of fundamental physical mechanisms that can inform the neural network. For that, we first define the following residuals of conversation equations of momentum, mass, and energy (corresponding to Eqs.(\ref{eq:U_momentum_equ}-\ref{eq:Temp_equ})) as
\begin{align}
    \label{eq:res}
    \left\{
           \begin{array}{ll}
         &\bm{r}_M: =  \rho ( {\bm{u}}_{,t} +  \bm{u}\cdot \nabla \textbf{u}-\bm{g}) + \nabla p - 2\mu \Delta \bm{u} \\
         &{r}_C: = \nabla\cdot \bm{u}  \\
         & {r}_T: =  (\rho c_p T)_{,t} +  \bm{u}\cdot \nabla (\rho c_p  T) + (\rho  L f_L)_{, t} + \bm{u}\cdot (\rho L \nabla f_L) - \kappa \nabla^2 T - Q_T  
            \end{array}
            \right.
\end{align}
%were $\bm{r}_M$, $\bm{r}_C$, and $\bm{r}_T$ defined corresponding to
% Then, the residuals of these PDEs ($\bm{r}_M$, ${r}_C$, ${r}_T$) are penalized in the loss function to represent. 
 
 To have well-posed systems, appropriate initial and boundary conditions are often necessary. In this paper, the initial boundary conditions are treated as part of the labeled data constraint. For boundary conditions (BCs), many existing PINN  frameworks utilize a ``soft" approach by designing additional loss components defined on the collocation points of boundaries to constrain the BCs. The downsides of this approach are two-fold: (1) The accuracy of satisfying the BCs is not guaranteed ; (2) The assigned weight of BC loss can affect learning efficiency, and no theory is existed to guide choosing the weight at this point.
 
 In this paper, we treat the Dirichlet BC in a ``hard" way by using a particular portion of the neural network to purely satisfy the prescribed Dirichlet BC. For that, we first define a Heaviside function as
\begin{equation} \label{eq: H_regularized}
H_{\epsilon}(\bm{x})=\left\{
\begin{aligned}
& 1- cos [d(\bm{x}) \pi/\epsilon]  & \text{if} \quad d(\bm{x}) &< &\epsilon \\
&{  \quad \quad \quad \quad  }1 & \quad \text{ if} \quad d(\bm{x}) \  &\geq  &\epsilon
\end{aligned}
\right.
\end{equation}
where $d (\bm{x})$ is the distance to the Dirichlet boundary. $\epsilon$ defines a artificial thickness of the boundary. With $H_{\epsilon}(\bm{x})$, the predictions of the neural networks are defined as
\begin{align}
\label{bc__u}
&\bm{u}_{NN} = \bm{u}_{bc}[1-H_{\epsilon}(\bm{x})] + \bm{u}H_{\epsilon}(\bm{x})
\\
\label{bc__p}
&p_{NN} = p_{bc}[1-H_{\epsilon}(\bm{x})] + p H_{\epsilon}(\bm{x})
\\
\label{bc__T}
&T_{NN} = T_{bc}[1-H_{\epsilon}(\bm{x})] + T H_{\epsilon}(\bm{x})
\end{align}
where $\bm{u}_{bc}$, $p_{bc}$, and $T_{bc}$ are the prescribed velocity, pressure, and temperature. $\bm{u}$, $p$, and $T$ are the solutions that satisfy the PDEs in the interior. Since $H_{\epsilon}(\bm{x})$ smoothly changes from 1 to 0 as $d (\bm{x})$ approaches to 0, the prediction will automatically satisfy the prescribed values by definition, without needing additional constraint. Then, the loss term from the PDEs with embedded Dirichlet BCs is defined as. 
\begin{align}
 \label{Loss_pde1}
 L_{pde}^{1}(\bm{W}, \bm{b})= \frac{1}{N_{r1}} \sum_{i=1}^{N_{r1}} \bm{r}_M [{\bm{u}_{NN}} (\bm{x}_i, t_i, \bm{W}, \bm{b}), {p}_{NN} (\bm{x}_i, t_i, \bm{W}, \bm{b}), {T}_{NN} (\bm{x}_i, t_i, \bm{W}, \bm{b}))]^2 \\\nonumber
 + \frac{1}{N_{r1}} \sum_{i=1}^{N_{r1}}  {r}_C [{\bm{u}_{NN}} (\bm{x}_i, t_i, \bm{W}, \bm{b}), {p}_{NN}(\bm{x}_i, t_i, \bm{W}, \bm{b}), {T}_{NN} (\bm{x}_i, t_i, \bm{W}, \bm{b}))]^2 \\\nonumber
  + \frac{1}{N_{r1}} \sum_{i=1}^{N_{r1}} {r}_T [{\bm{u}_{NN}} (\bm{x}_i, t_i, \bm{W}, \bm{b}), {p}_{NN} (\bm{x}_i, t_i, \bm{W}, \bm{b}), {T}_{NN} (\bm{x}_i, t_i, \bm{W}, \bm{b}))]^2
\end{align}
where $N_{r1}$ denotes the number of collocation points to constrain the PDEs.

In the thermal-fluid flow model for metal AM processes, Neumann BCs incorporate surface tension for flow field and laser for thermal field. In this paper, the Neumann BCs are handled by the conventional way, where the following term is added in the loss function
\begin{align}
 \label{Loss_pde2}
 L_{pde}^{2}(\bm{W}, \bm{b})=   \frac{1}{N_{r2}} \sum_{i=1}^{N_{r2}}\left\{2\mu\nabla^S{\bm{u}_{NN}} (\bm{x}_i, t_i, \bm{W}, \bm{b})\cdot \bm{n} - {p}_{NN} (\bm{x}_i, t_i, \bm{W}, \bm{b}) \bm{n}- \bm{\tau}(\bm{x}_i, t_i)\right\}^2\\\nonumber
 + \frac{1}{N_{r3}} \sum_{i=1}^{N_{r3}} [\kappa \nabla {{T}_{NN}} (\bm{x}_i, t_i, \bm{W}, \bm{b})  \cdot \bm{n}- {q}(\bm{x}_i, t_i)]^2 
\end{align}
where $N_{r2}$ and $N_{r3}$ denote the number of collocation points on the fluid and temperature Neumann boundaries, respectively. 

%where $N_{r1}$ is the number of collocation points in the interior. $N_{r2}$ and $N_{r3}$ are the number of collocation points of velocity and temperature on the Neumann boundaries.

With above definitions, The hidden parameters $\bm{W}$ and $\bm{b}$ are obtained by minimizing the following total loss function, which are a linear combination of data constraint of $L_{data}$ and PDE constraints of $L_{pde}^{1}$ and $L_{pde}^{2}$.

\begin{align}
 \label{eq:Loss_functon}
&\displaystyle \min_{\bm{W},\bm{b}} L (\bm{W},\bm{b}) = (1-\lambda_{pde}^1-\lambda_{pde}^2) L_{data}(\bm{W},\bm{b}) + \lambda_{pde}^1 L_{pde}^{1}(\bm{W},\bm{b})+ \lambda_{pde}^2 L_{pde}^{2}(\bm{W},\bm{b})
\end{align}
 $\lambda_{pde}^1$ and  $\lambda_{pde}^2$ are two positive numbers between 0 and 1, which define the weight of the physical law constraints in the loss function.

\subsubsection{Learning procedure}
The PINN model is trained by minimizing the loss function defined in Eq.~\ref{eq:Loss_functon}. Most of current machine learning frameworks solve the optimization problem by a stochastic gradient
descent (SGD) algorithm, which is a stochastic approximation of the gradient descent (GD) optimization~\cite{ruder2016overview}. SGD only uses a subset of collocation points, randomly sampled from the input space at each iteration, to calculate the directional gradient. Research shows that SGD works very well to skip bad local minima. One issue with SGD is the oscillation of gradient direction caused by the random selection of sampled collocation points. In this paper, the Adam method~\cite{kingma2014adam} that combines adaptive learning rate and momentum methods is used to improve convergence speed~\cite{kingma2014adam}.

The PINN learning process needs the spatial and temporal derivatives of $\bm{u}$, $p$, and $T$, which can be accurately and efficiently calculated by using automatic differentiation (AD)~\cite{AD}. The basic idea of AD is to use the chain rule to back-propagate derivatives from the output layer to the input layer since the connection between layers of a neural network is analytically defined. Compared to numerical differentiation techniques (e.g., finite difference and finite element), AD does not suffer from truncation or round-off errors, resulting in much higher accuracy. AD has been gaining increasing attention in the machine learning community and has been implemented in many modern deep learning frameworks, such as TensorFlow~\cite{abadi2016tensorflow}, PyTorch~\cite{paszke2019pytorch}, Theano~\cite{bastien2012theano}, and Caffe~\cite{jia2014caffe}. In this paper, the PINN formulation is implemented in TensorFlow.

\section{Applications}\label{sec: Applications}
\subsection{Solidification of aluminum in a graphite mold without labeled data}

 \begin{figure}[!htbp]
\centering
\includegraphics[scale=0.25]{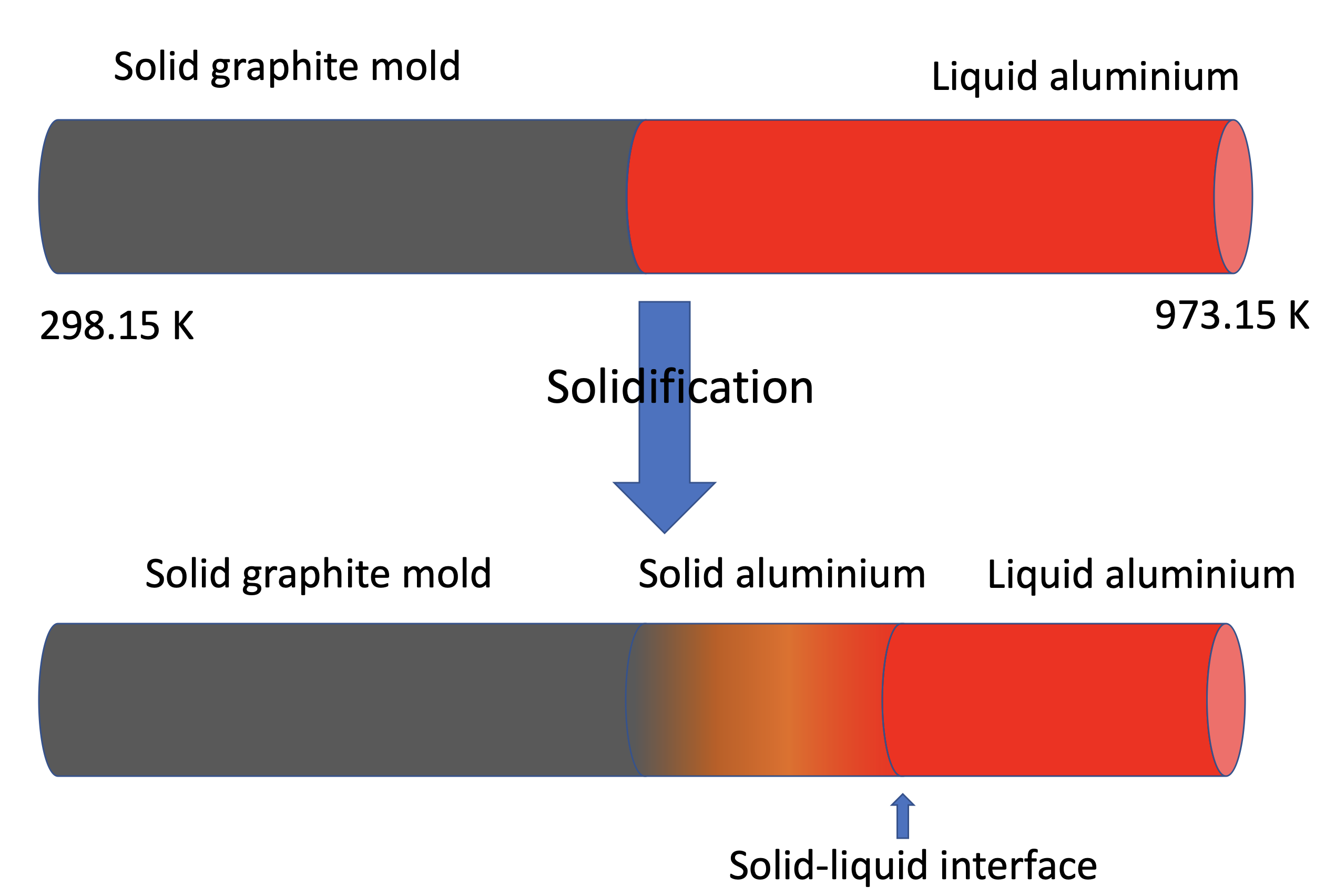}
\caption{1D solidificaton process from~\cite{dantzig2016solidification}}
\label{fig:solidification}
\end{figure}

 The solidification process of aluminum in a graphite mold from the textbook \textit{Solidification} by Dantzig and Rappaz~\cite{dantzig2016solidification} is investigated to assess the performance of the PINN formulation. Only thermodynamics with phase transition is considered in the simulations here. Figure~\ref{fig:solidification} shows the problem setup, where the left half of the domain (-0.4 m $ \leq x \leq $ 0.0 m) is occupied by a solid graphite mold with temperature $T_{low}$ = 298.15 K, and the right half of the domain (0.0 m $< x \leq $ 0.4 m) is occupied with liquid aluminum with temperature $T_{high}$ = 973.15 K, which is higher than the melting temperature of aluminum $T_{melt}=933.15K$. As depicted in Figure~\ref{fig:solidification} (lower), solidification occurs by transferring heat from the aluminum into the mold, and the solid-liquid interface propagates towards the right end. The material properties of the graphite mold and aluminum are given in Table~\ref{tab:solid_mat}. The analytical solutions have been derived in~\cite{dantzig2016solidification} for this problem and are specified as follows.
 
 \begin{table}[ht!]
 \centering
 \caption{\label{tab:solid_mat}Definition of material properties for the solidification problem}
 \begin{tabular}{llll}
  \toprule
  Materials & Graphite & Aluminum (solid) & Aluminum (liquid) \\\hline

  Density ($kg/m^3$) & 2200 & 2555 & 2555 \\

  Specific heat ($J/(kg\ K)$) & 1700 & 1190 & 1190 \\

  Thermal conductivity ($W/(kg\ K)$) & 100 & 211 & 91 \\
  Latent heat ($J/kg$) & - & $398000$ & - \\\hline

 \end{tabular}
\end{table}
\begin{align}
 &x^* = 7.095 \times 10^{-3}\sqrt{t} \ s\\
 &T_m= 769.95+471.8erf(\frac{96.69x}{\sqrt{t}}) \ K\\
 &T_s= 769.95+360.2erf(\frac{60.02x}{\sqrt{t}}) \ K\\
 &T_l= 973.15-111.4erfc(\frac{91.39x}{\sqrt{t}})\ K
\end{align}
where $x^*$ is the location of solid-liquid interface over time, $T_m$, $T_s$, and $T_l$ are the temperature distribution in the mold, solid aluminum, and liquid aluminum, respectively.

Despite the simplicity, solving this problem provides valuable insights into the solidification behavior and the performance of the machine learning model. The PINN model employs 5 hidden layers and 200 neurons of each layer, which provides better results over others based on our non-exhaustive investigation. The neural network is physically informed with the energy conservation principle defined in Eq.~\ref{eq:Temp_equ} and trained without labeled data-set. The PINN model predicts the temperature distribution from $t$ = 5 s to $t$ = 10 s.
 
\begin{figure}[!htbp]
\centering
\includegraphics[scale=0.55]{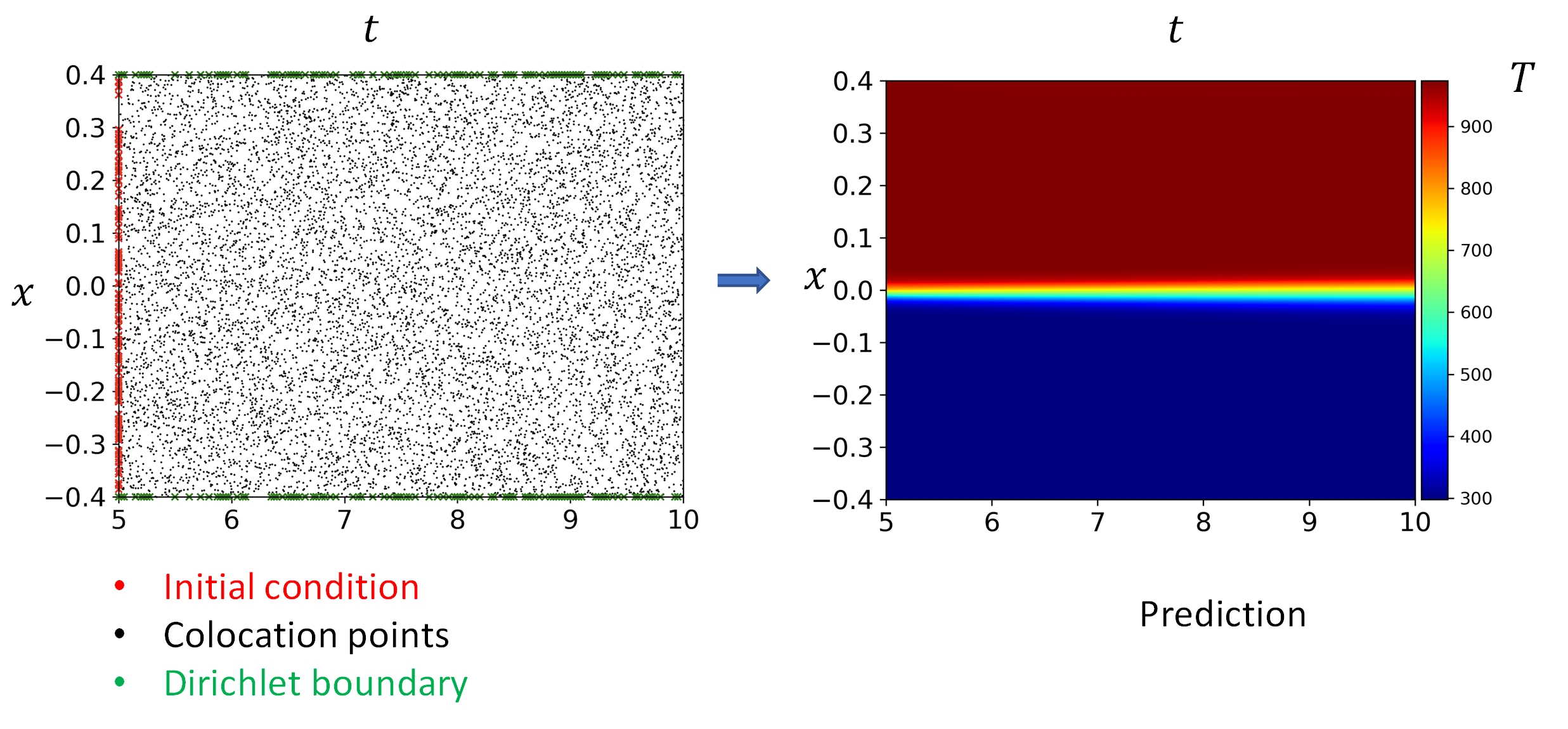}
\caption{\textcolor{black}{PINN model for the solidification problem. Left: PINN setup. Right: Temperature prediction.}}
\label{fig:PINN_tem_2D}
\end{figure}

\begin{figure}[!htbp]
	\centering
{\includegraphics[width=4in]{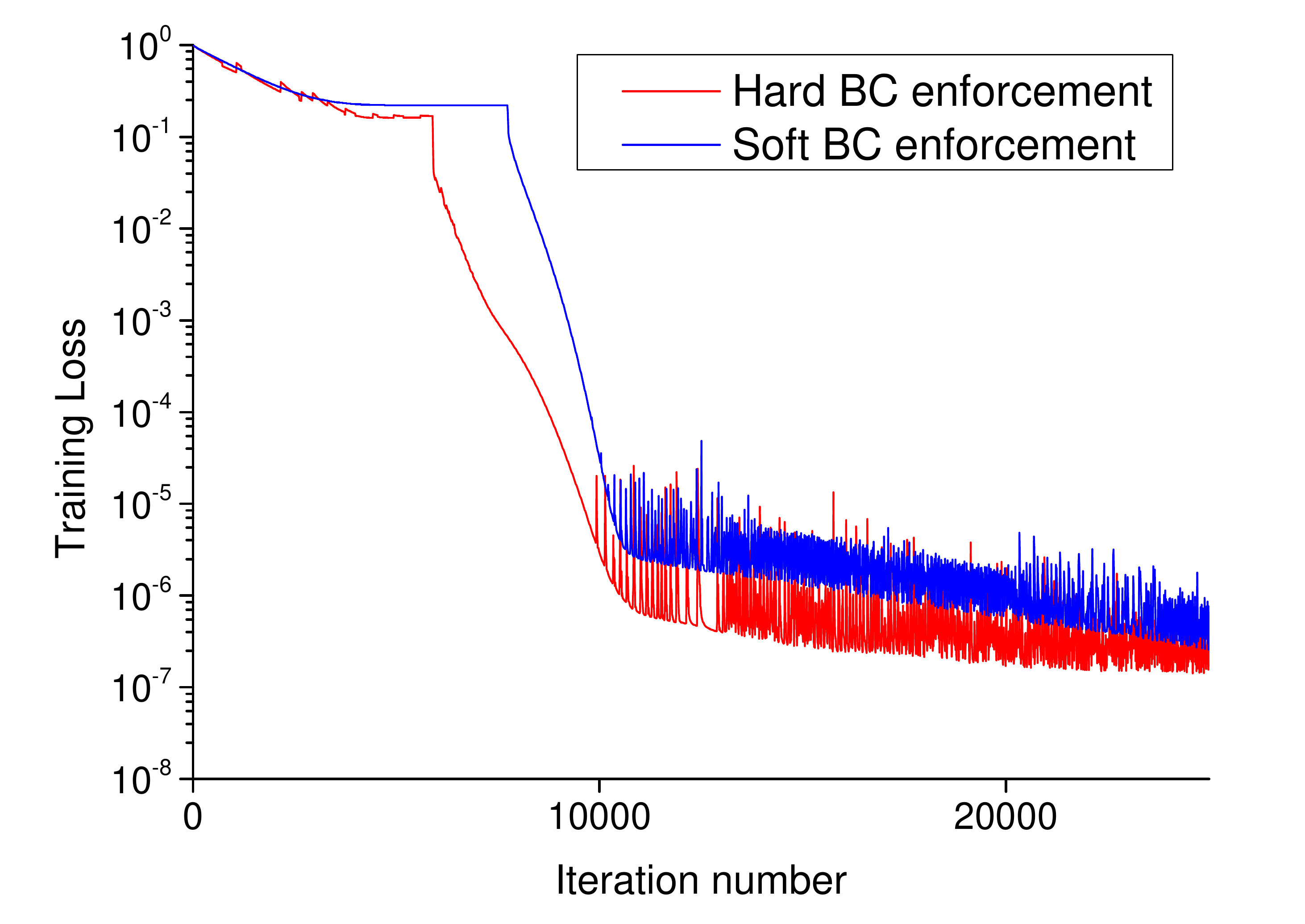}}
{\includegraphics[width=4in]{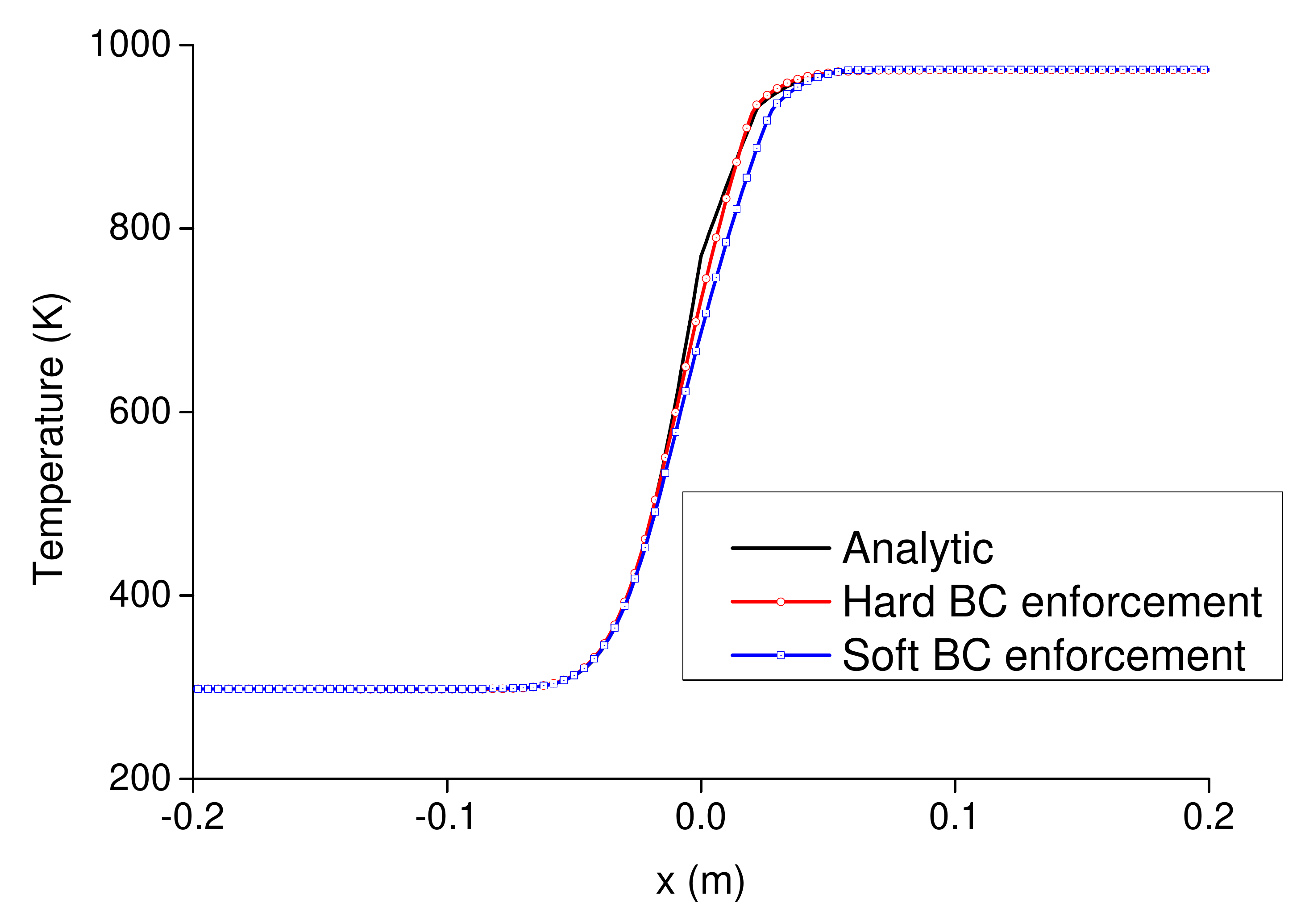}}\\	
	\caption{\textcolor{black}{Comparison between ``hard" BC and ``soft" BC. Left: learning speed. Right: The temperature distribution at 10 s.}}
	\label{fig:BC_comp}
\end{figure}

Figure~\ref{fig:PINN_tem_2D} illustrates the PINN setup and the resulting temperature prediction in the space-time (${x}-t$) slab for the solidification process. We compare the performance of the proposed ``hard" approach with the conventional ``soft" approach for the Dirichlet boundary condition in Figure~\ref{fig:BC_comp}, which depicts the learning process and temperature predictions at 10 s. The plot shows that the ``hard" approach can not only facilitate the learning process (see Figure~\ref{fig:BC_comp} (left)) but also produce more accurate temperature prediction (see Figure~\ref{fig:BC_comp} (right)).
%he plot shows the accuracy difference between both approaches is non-noticeable, but the ``hard" approach can obviously facilitate the learning process (See Figure~\ref{fig:BC_comp} (left)).

\begin{figure}[!htbp]
	\centering
{\includegraphics[width=4in]{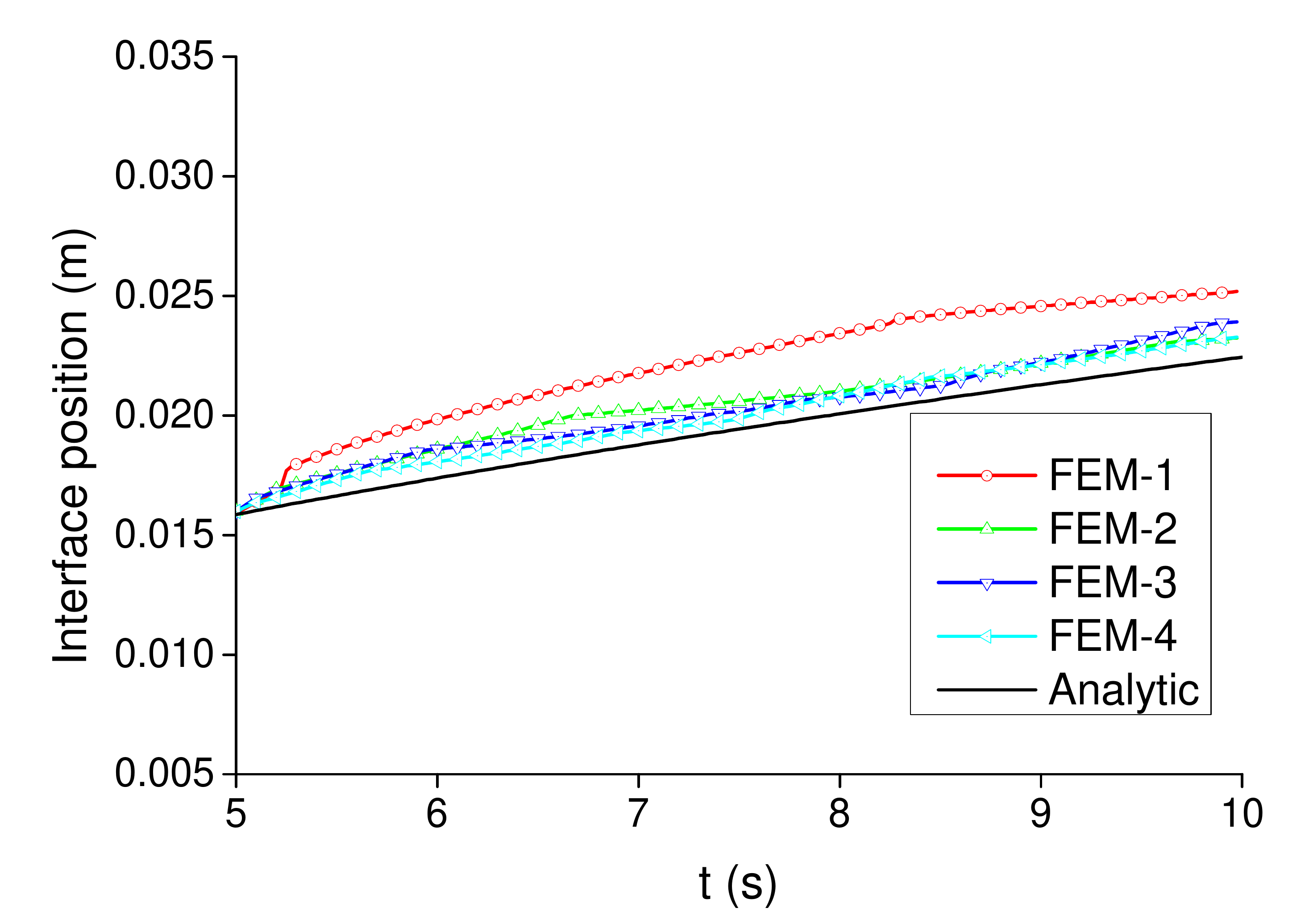}}
{\includegraphics[width=4in]{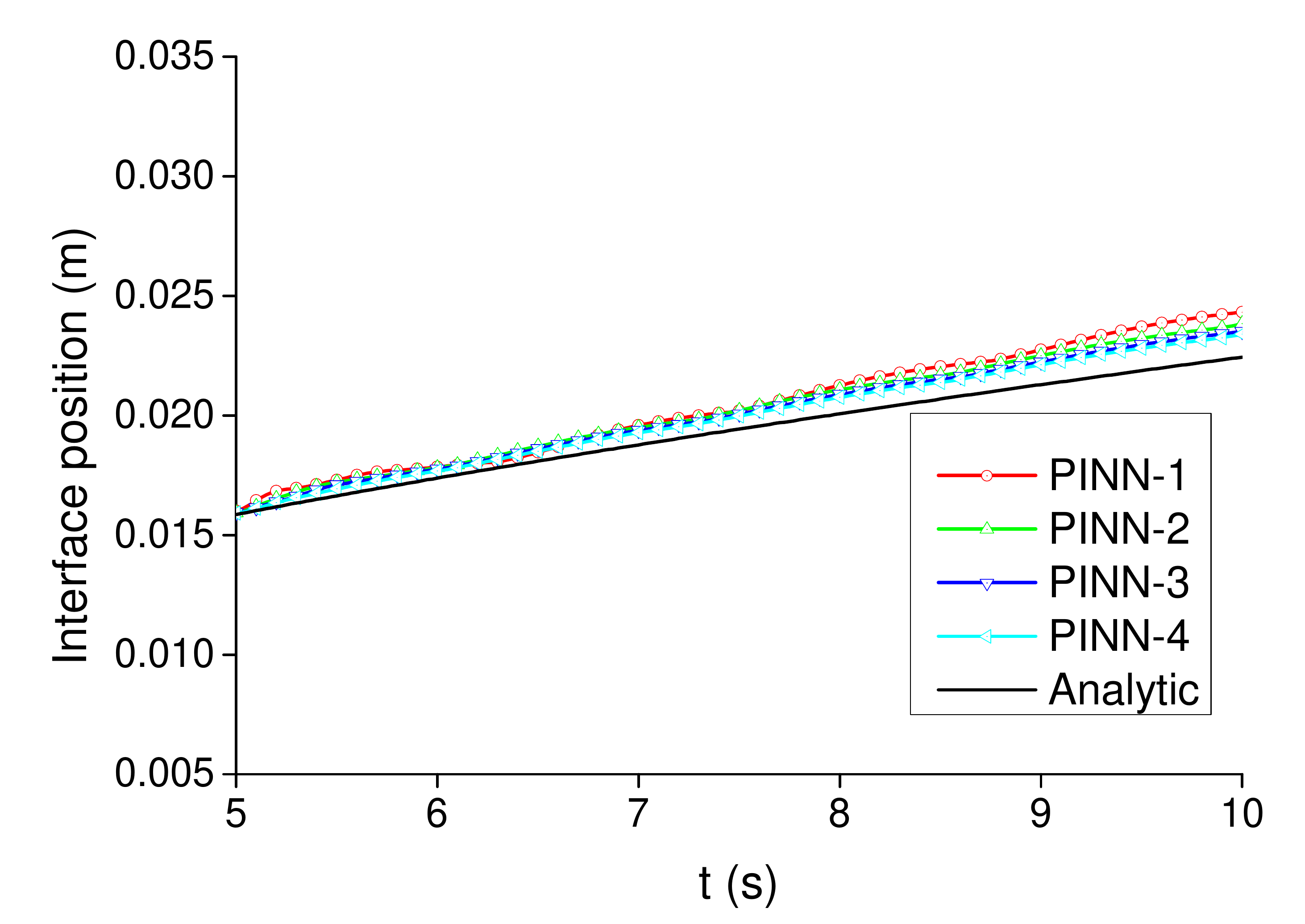}}
	\caption{\textcolor{black}{Predicted time history of solid-liquid interface position. Left: refinement study of FEM. Right: refinement study of PINN.}}
	\label{fig:fem_PINN_comp}
\end{figure}

\begin{figure}[!htbp]
\centering
\includegraphics[scale=0.2]{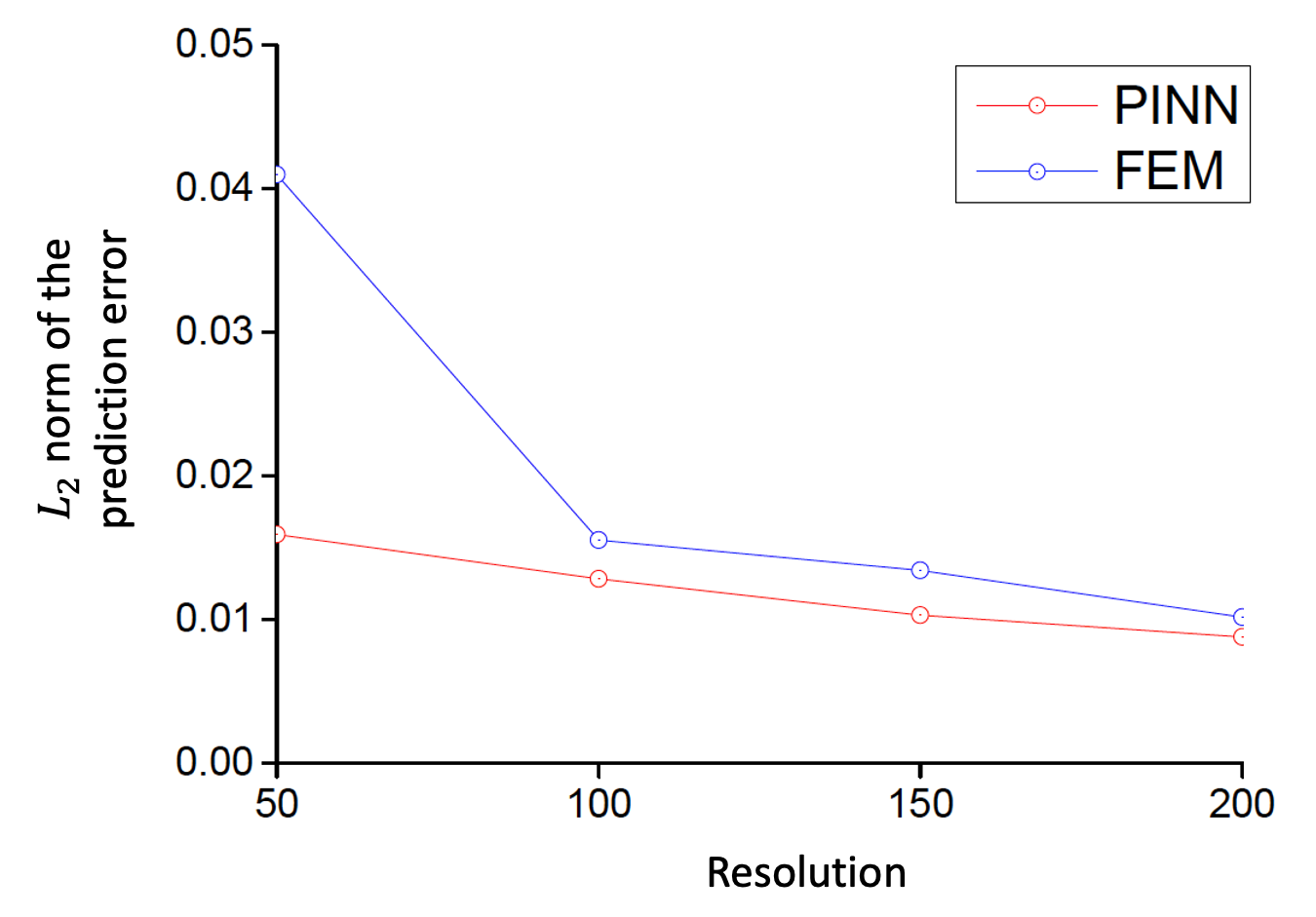}
\caption{$L_2$ norm of prediction error of PINN and FEM with different resolutions.}
\label{fig:refinement}
\end{figure}

One important question is how does the PINN's predictive capability compare with traditional numerical methods, such as finite element method (FEM). To answer this question, we simulate this solidification problem by using PINN with four different numbers of collocation points and linear FEM with four equivalent resolutions (\textcolor{black}{$N_{x}$= 50, 100, 150, and 200} along $x$ direction). Figure~\ref{fig:fem_PINN_comp} shows the predictions of PINN and FEM for the time history of solid-liquid interface position with the four resolutions. The convergence rate of error of temperature prediction over the ${x}-t$ slab is shown in Figure~\ref{fig:refinement}. The two plots indicate that PINN and FEM obtain similar convergence rates. However, when the resolution is low, the PINN still attains high accuracy while noticeable discrepancy is observed for the standard FEM.  

%\section{3D Thermal-fluid case}
\subsection{NIST AM-bench test series}
In this section, we apply the PINN framework to the Additive Manufacturing Benchmark (AM-Bench) test series conducted by the National Institute of Standards and Technology (NIST)~\cite{AM_bench, AM_bench1}. In 2018, NIST performed a series of metal AM experiments with different manufacturing parameters, which attracted blind simulations to compare with the in-situ and ex-situ measurements, such as temperature, melt pool dimensions and micro-structures~\cite{AM_bench}. The archived experimental measurements provide valuable benchmark data for modelers to test the predictive capabilities of simulation models. In this paper, we use the proposed PINN framework to predict the temperature, melt pool fluid dynamics, melt pool dimensions, and cooling rates during the NIST AM processes, which corresponds to \textcolor{black}{the first challenge in the NIST AM-Bench test series~\cite{AM_bench1}. To the best knowledge of the authors, this is the first application of PINN to three dimensional metal AM processes modeling.}

% NIST selective laser beam melting processes of pure Inconel 625 (IN625) substrates with three different laser power and scanning speed.
 
\begin{figure}[!htbp]
\centering
\includegraphics[scale=0.4]{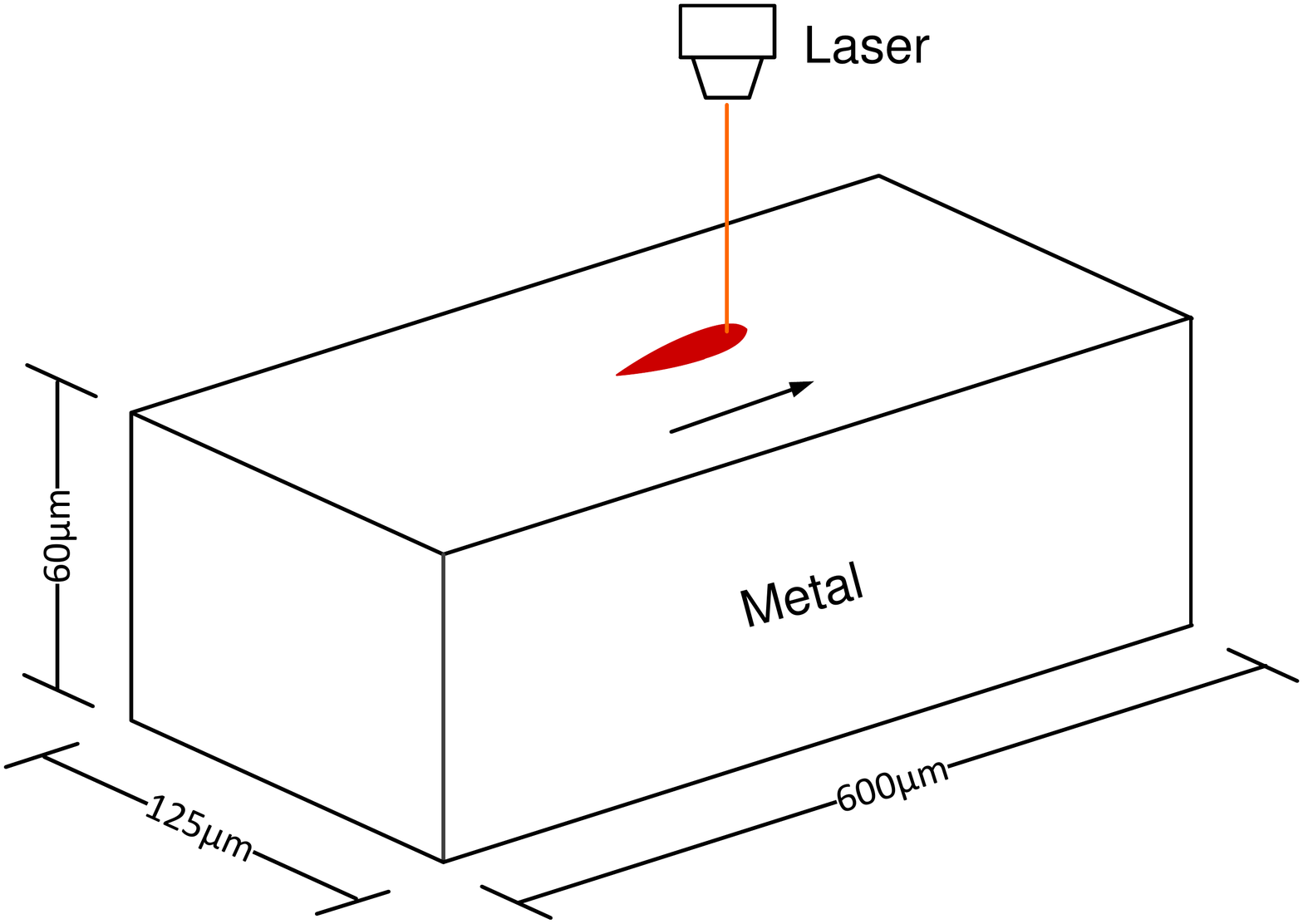}
\caption{NIST AM-bench test series}
\label{fig:solid_diag_1}
\end{figure}

Figure~\ref{fig:solid_diag_1} shows a schematic description of the NIST AM-Bench test, a selective laser beam melting process of a pure Inconel 625 (IN625) substrate. The material properties used in this paper are summarized in Table~\ref{tab:IN625_mat}. The laser is applied by imposing the following moving heat flux on the substrate.
\begin{align}
\kappa \nabla{T}\cdot \bm{n} = q_{laser} = \frac{2Q\eta}{\pi r_b^2}exp(\frac{-2((x-V_st)^2+y^2)}{r_b^2})
\end{align}
where $Q$ is the laser power, $\eta$ is the absorptivity, $r_b$ is the laser beam radius, $V_s$ is the laser scanning speed. \textcolor{black}{$\eta = 0.43 $ and $r_b = 50$ $\mu$m are used in this paper.} Table~\ref{tab:threecases} lists the laser power and scanning speed of three cases used in the NIST experiments. Based on the fact that the top surface deformation is relatively small compared with the melt pool dimension, a flat top surface is assumed, and the following boundary condition is applied for the fluid field.
\begin{align}
- p\bm{n} + 2\mu \nabla^{s}\bm{u}\cdot \bm{n} = \bm{\tau} = \frac{d\sigma}{dT} [\nabla T - (\nabla T\cdot \bm{n})\bm{n}]
\end{align}
where $\frac{d\sigma}{dT}$ is the Marangoni coefficient, which is only effective in the tangential direction of the temperature gradient. Except for the top surface, no-slip and fixed reference temperature boundary conditions are used for other surfaces.

The PINN model employs a fully connected neural network with 5 hidden layers and 250 neurons per layer, which provides a better balance of accuracy and training efficiency over others based on our non-exhaustive investigation. The model is physically informed by the conservation laws of momentum, mass, and energy defined in Eqs.~\ref{eq:U_momentum_equ}-~\ref{eq:Temp_equ}. A total of $788,651$ collocation points in the $\bm{x}-t$ slab are used to penalize the PDE residuals (see Eq.~\ref{eq:res}) in the loss function. Due to the limited point-wise measurements provided by NIST, we use high-fidelity finite element thermal-fluid simulations to generate the labeled data-sets to facilitate the training. The simulated results are treated as ``ground truth" after being compared with the NIST experimental measurements. \textbf{Appendix} briefly presents the core formulation of the finite element formulation that has also been validated for other metal manufacturing process modeling~\cite{yan2018fully}. The setup for FEM simulations, such as time step and mesh size, is also specified in \textbf{Appendix}. The FEM simulations have been run up to 2.0 ms for all the three cases listed in Table~\ref{tab:threecases}, but only a small portion of the simulated data between 1.2 ms and 1.5 ms is used as labeled training data in the PINN model, which then predicts the manufacturing processes for a wider time interval from 0 ms to 2.0 ms. 

\begin{table}[!htbp]
 \centering
 \caption{\label{tab:IN625_mat} Mechanical properties of IN625}
 \begin{tabular}{llll}
  \toprule
  Name & Property & Value  \\\hline
  Density & $\rho (kg\cdot m^{-3})$ & $8440$ \\
  Solidus temperature & $T_s(K)$ & $1563$   \\
  Liquid temperature & $T_l(K)$ & $1623$   \\
  Solid specific heat capacity & $c_{ps}(J\cdot kg^{-1}\cdot K^{-1})$ & $0.2441T+338.39$   \\
  Liquid specific heat capacity & $c_{pl}(J\cdot kg^{-1}\cdot K^{-1})$ & $709.25$  \\
  Solid solid conductivity & $\kappa_{s}(W\cdot m^{-1}\cdot K^{-1})$ & $3.0\times 10^{-5}T^2-0.0366T+18.588$  \\
  Liquid solid conductivity & $\kappa_{l}(W\cdot m^{-1}\cdot K^{-1})$ & $30.078$   \\
  Latent heat of fusion & $c_L(KJ\cdot kg^{-1}\cdot K^{-1})$ & $290$   \\
  Dynamic viscosity & $\mu (Pa\cdot s)$ & $7\times 10^{-3}$   \\
  Marangoni coefficient & $\frac{\partial \gamma}{\partial T} (N\cdot m^{-1} \cdot K^{-1})$ & $-2\times 10^{-5}$  \\
  Reference temperature & $T_{ref} (K)$ & $295$ \\\hline
 \end{tabular}
\end{table}

\begin{table}[!htbp]
 \centering
 \caption{\label{tab:threecases} Three laser parameters}
 \begin{tabular}{llll}
  \toprule
  Parameters & Case A & Case B & Case C  \\\hline
  Laser Power & 150 W & 195 W & 195 W  \\
  Scan Speed & 0.4 m/s & 0.8 m/s & 1.2 m/s \\\hline
 \end{tabular}
\end{table}

\begin{figure}[!htbp]
\centering
\includegraphics[scale=0.6]{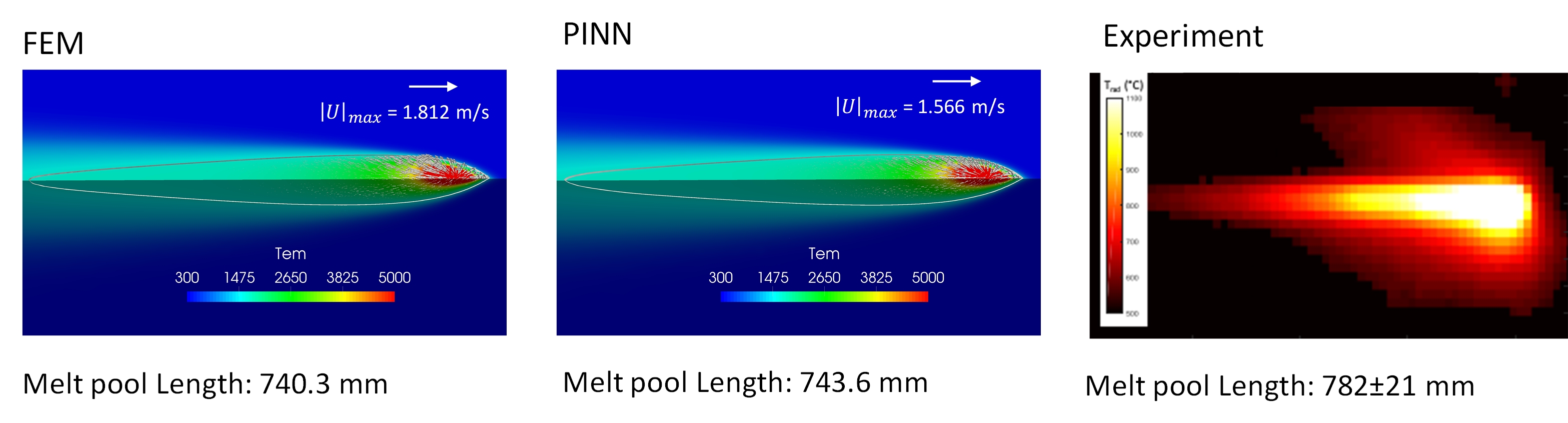}
\caption{\textcolor{black}{Comparison of the predictions of the temperature and melt pool fluid dynamics of FEM, PINN and experiment for case B (195 W, 0.8 m/s) at quasi-steady state (2 ms), when the melt pool shape is not changing. Left: FEM prediction. Middle: PINN prediction. Right: Thermal video frame based on radiance temperature.}}
\label{fig:solid_diag_fem}
\end{figure}

\begin{figure}[!htbp]
\centering
\includegraphics[scale=0.4]{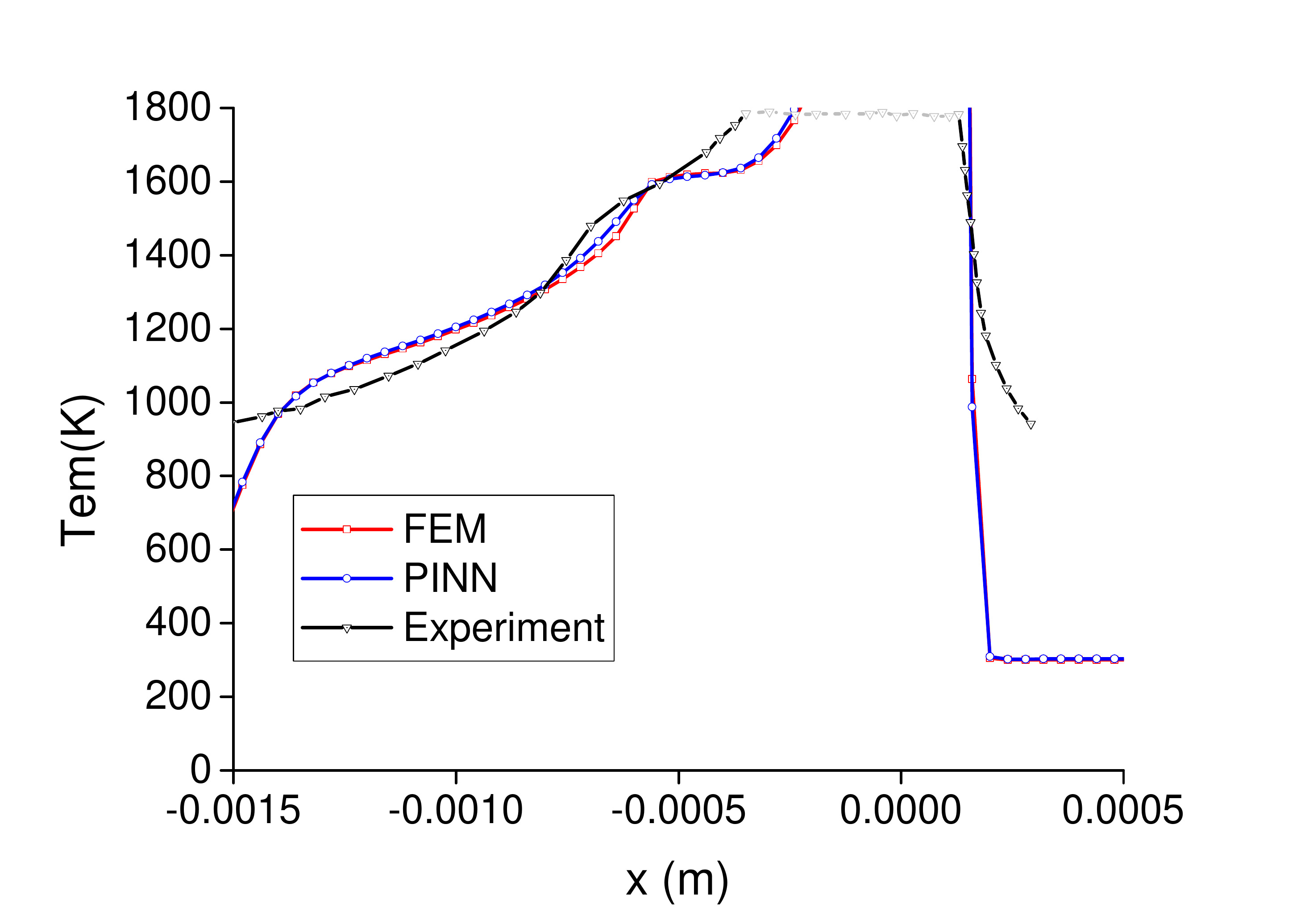}
\caption{Temperature profile along the scan track for case B (195 W, 0.8 m/s) at quasi-steady state. Experimental data extracted from NIST AM-Bench test series~\cite{heigel2018measurement} are also plotted for comparison.}
\label{fig:centerline_fem_pinn}
\end{figure}

We first compare the predicted results of FEM and PINN with available experimental data for case B (195 W, 0.8 m/s). The purpose is two-fold: 1) Ensure the credibility of FEM data as the training data; 2) Validate the PINN model. Figure~\ref{fig:solid_diag_fem} shows the temperature field, melt pool fluid dynamics, and melt pool shape at 2.0 ms. The fast-moving laser, along with the effect of a negative Marangoni coefficient that drives the liquid metal from higher temperature to lower temperature, leading to a long and shallow melt pool. The predicted temperature profile along the scan track and experimental measurement extracted from~\cite{heigel2018measurement} is plotted in Figure~\ref{fig:centerline_fem_pinn} for comparison. The predicted results by both FEM and PINN show good agreement with available experimental data. Figure~\ref{fig:solid_diag_fem} and Figure~\ref{fig:centerline_fem_pinn} also show that the PINN model, with a moderate amount of training data, can generate very similar predictions of temperature, melt pool length, and melt pool fluid velocity to those of FEM.

\begin{figure}[!htbp]
\centering
\includegraphics[scale=0.6]{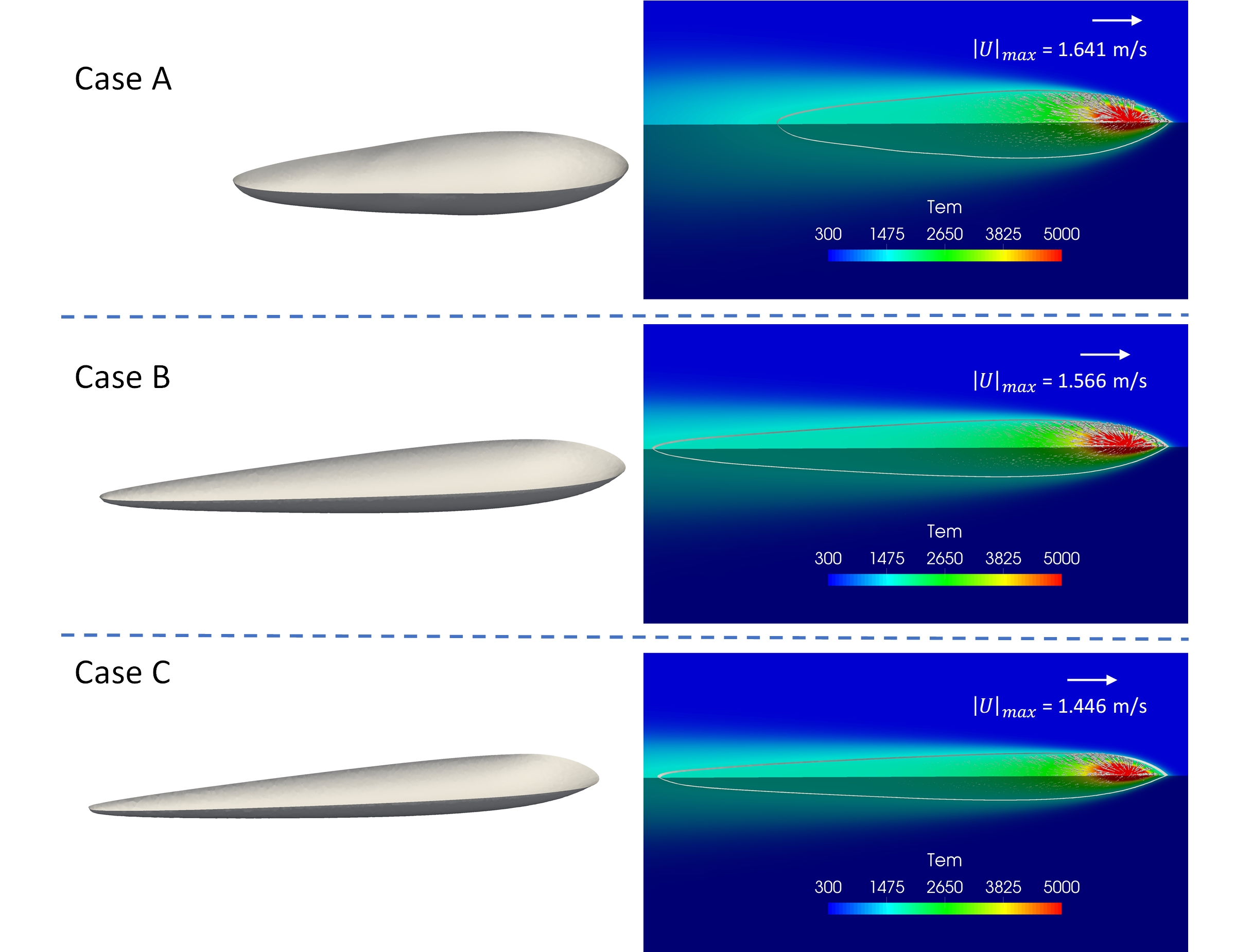}
\caption{\textcolor{black}{Melt pool shape and melt pool flow velocity predicted by PINN for case A, B and C at quasi-steady state (2 ms).}}
\label{fig:all_cases_shapes}
\end{figure}

\begin{table}[!htbp]
 \centering
 \caption{\label{tab:three_melt_shape}Melt pool dimensions of case A, B and C.}
 \begin{tabular}{lllll}
  \toprule
 Cases &Approaches & Length ($\mu m$) & Width ($\mu m$) & Depth ($\mu m$)  \\\hline
 Case A & PINN & 594.8 \textcolor{black}{(9.7\%)} & 193.3 & 64.0  \\
 & FEM & 584.4 \textcolor{black}{(11.3\%)} & 190.8 & 62.8\\
 & \textcolor{black}{Gan et al.~\cite{gan2019benchmark}} & \textcolor{black}{542 (17.8\%)} & - & -\\
 & Experiment~\cite{heigel2018measurement} & $659\pm 21$ & - & -  \\\hline
 Case B &  PINN & 740.3 \textcolor{black}{(5.1\%)} & 160.0 & 52.8  \\
&  FEM & 743.6 \textcolor{black}{(4.7\%)} & 157.5 & 52.5 \\
& \textcolor{black}{Gan et al.~\cite{gan2019benchmark}} & \textcolor{black}{843 (8.1\%)} & - & -\\
&  Experiment~\cite{heigel2018measurement} & $782\pm 21$ & - & - \\\hline
 Case C &  PINN & 732.5 \textcolor{black}{(2.9\%)} & 131.5 & 43.2  \\
& FEM & 727.2 \textcolor{black}{(3.6\%)} & 130.3 & 42.6 \\
& \textcolor{black}{Gan et al.~\cite{gan2019benchmark}} & \textcolor{black}{785 (4.1\%)} & - & -\\
&  Experiment~\cite{heigel2018measurement}  & $754\pm 46$ & - & -  \\
  \bottomrule
 \end{tabular}
\end{table}

We then apply the PINN model to all the three cases listed in Table~\ref{tab:threecases}. The predicted melt pool shape and the fluid velocity field within the melt pool are presented in Figure~\ref{fig:all_cases_shapes}. The laser results in sufficiently high velocity in the melt pool, which reaches up to 1.641 m/s, 1.566 m/s, and 1.446 m/s for the case A, B, and C, respectively. The predicted melt pool dimensions compared with the present FEM results, \textcolor{black}{the thermal-fluid simulation results by Gan et al.~\cite{gan2019benchmark} that won an award in the NIST AM-bench competition}, and available NIST experimental measurements are listed in Table~\ref{tab:three_melt_shape}. \textcolor{black}{For melt pool length, the relative discrepancy (with respect to the mean NIST experimental measurements) of PINN, FEM and Gan's predictions are 9.7 $\%$, 11.3 $\%$ and 17.8 $\%$ for case A, 5.1 $\%$, 4.7  $\%$  and 8.1 $\%$ for case B, and 2.9 $\%$, 3.6 $\%$ and 4.1 $\%$ for case C.} Similar accuracy is observed among these approaches. \textcolor{black}{Based on the model predictions, we notice that case A generates the smallest melt pool length but the biggest width and depth, which could be due to the low scanning speed that gives the laser more time to melt the underneath metal. In contrast, case B and C have a slightly bigger laser power but much faster scanning speeds, which consequently result in longer melt pool lengths but smaller widths and depths. Between case B and case C with the same laser power, lower spanning (case B) speed results in bigger melt pool dimensions in all directions.} 

A critical factor in metal AM is the cooling rate, which profoundly influences dendrite arm spacing, grain structure, micro-segregation, and hot cracking. In this paper, the cooling rate is calculated as

\begin{align}
 \label{eq:cooling_rate}
& R_c = \frac{T_s-1273.15 K}{t_c}
\end{align}
where $t_c = ({D_s- D_{1273.15}})/{V_s}$, the cooling time interval determined by dividing the distance between solidus temperature and 1273.15 K by the scanning speed $V_s$.

\begin{table}[!htbp]
 \centering
 \caption{\label{tab:melt_shape} Cooling rates of case A, B and C.}
 \begin{tabular}{lll}
  \toprule
 Cases &Approaches  & solid cooling rate ($K/s$) \\\hline
 Case A & PINN &  $8.54\times 10^5$ \textcolor{black}{(37.7\%)} \\
 & FEM  & $8.00\times 10^5$ \textcolor{black}{(29.0\%)} \\
 & \textcolor{black}{Gan et al.~\cite{gan2019benchmark}} & \textcolor{black}{$5.11\times 10^5$ (17.6\%)} \\
 & Experiment~\cite{heigel2018measurement} & $6.20\times 10^5 \pm 7.99\times 10^4$ \\\hline
 Case B &  PINN & $8.59\times 10^5$ \textcolor{black}{(8.1\%)} \\
& FEM & $8.16\times 10^5$ \textcolor{black}{(12.7\%)} \\
& \textcolor{black}{Gan et al.~\cite{gan2019benchmark}} & \textcolor{black}{$6.89\times 10^5$ (26.3\%)} \\
&  Experiment~\cite{heigel2018measurement}  & $9.35\times 10^5 \pm 1.43\times 10^5$\\\hline
 Case C &  PINN & $1.38\times 10^6$ \textcolor{black}{(7.8\%)} \\
& FEM  & $1.38\times 10^6$ \textcolor{black}{(7.8\%)} \\
& \textcolor{black}{Gan et al.~\cite{gan2019benchmark}} & \textcolor{black}{$11.30\times 10^5$ (11.7\%)} \\
&  Experiment~\cite{heigel2018measurement}   & $1.28\times 10^6 \pm 3.94\times 10^5$ \\
  \bottomrule
 \end{tabular}
\end{table}

Table~\ref{tab:melt_shape} presents the PINN prediction of cooling rate $R_c$ for the case A, B, and C. The results of FEM, Gan's results and experimental measurements are also listed for comparison. Both modeling and experiment show that cooling rate increases from case A to case C. \textcolor{black}{For the cooling rate, the relative discrepancy (with respect to the mean NIST measurements) of PINN, FEM and} \textcolor{black}{Gan's predictions are 37.7 $\%$, 29.0 $\%$ and 17.6 $\%$ for case A, 8.1 $\%$, 12.7 $\%$ and 26.3 $\%$ for case B, and 7.8 $\%$, 7.8 $\%$ and 11.7 $\%$ for case C.} Although all the models' prediction accuracy becomes lower compared with melt pool dimension prediction, we notice that the NIST measurements in cooling rates also exhibit significantly higher fluctuations than melt pool dimension measurements. If using this discrepancy as an accuracy metric, the proposed PINN model only underperforms in case A and outperforms in both case B and case C, which is advantageous compared with the other two high-fidelity FEM simulations that employ millions of elements.

%%-------------------------------------------------------------------------------------------------
%\section{Summary, conclusions and outlook}
%In this paper physical neural network is used to solve the multi-physics problem. The loss function consists the data loss and collocation loss in the current paper. The effects of the number of hidden layers, number of neurons per hidden layer, way of enforcement with Dirichlet boundary condition, and number of collocation points are analyzed. Three cases of Thermal-fluid problems in additive manufacturing are simulated to show the potentials of PINN in the multi-physics problem. The match between the FEM solution and the PINN solution shows the accuracy of our current method. 

\section{Conclusion}\label{sec:concl}

This paper presents the first attempt of using the physics-informed neural network (PINN) to predict the temperature and melt pool fluid dynamics in metal AM processes. We applied the PINN model to two representative metal manufacturing problems. The results show that the PINN can accurately predict the quantities of interest by only using a small amount of labeled training data. This paper is also the first few applications of scientific machine learning (SciML), currently confined to single-phase systems, to complex multi-scale and multi-physics problems that involve multi-phase fluid dynamics, heat transfer and phase transition. The two major technical contributions relevant to metal AM of the paper are:
\begin{itemize}
  \item A SciML framework for metal AM processes, which can accurately predict temperature, pressure, and velocity field without relying on big-data.
  \item A ``hard" approach for imposing Dirichlet boundary condition, which exactly imposes the prescribed value and speeds up the learning process.
\end{itemize}

Although deep learning models cannot replace conventional numerical tools that will continue to be the principal player, the initial success presented in this paper demonstrates PINN's potential on the modeling and prediction of complicated metal AM processes and paves the way for the broad adoption in advanced manufacturing.

We also have to admit that this paper does not comprehensively handle the complexity of metal AM processes. To be precise, the PINN model here does not resolve the ambient gas phase, free-surface deformation of the melt pool, and the evaporation phenomenon, although the effects are not crucial for the applications considered in the paper. In the future, the multi-phase Navier-Stokes will be enhanced with the evaporation model in the momentum equations, which was used in our control volume finite element model~\cite{Lin2020}, to capture the heat loss, composition change, and fluid motion induced by evaporation. Additional PDEs, such as convection equation of level set or volume-of-fluid used in the previous works~\cite{yan2018fully,wentao0}, will be incorporated into the PINN to enable modeling metal AM process at the powder scale.

\section{Acknowledgement}\label{sec:acknowledgement}

J. Yan is partially supported by ASME Robert M. and Mary Haythornthwaite Research Initiation Award and Singapore National Research Foundation (NRF2018-ITS004-0011). The PINN models were trained at the Texas Advanced Computing Center (Tacc) through a startup allocation on Frontera (CTS20014). These supports are greatly acknowledged.

\section{Appendix: High-fidelity FEM thermal-fluid simulations}\label{sec:Appendex}

The high-fidelity FEM thermal-fluid simulations make use of a residual-based variational multi-scale formulation (VMS) based on our previous work in~\cite{yan2018fully}. The core formulation is briefly presented as follows. Let $\mathcal{W}$ denote the testing function space for the Navier-Stokes and energy conversation equations, $\mathcal{V}$ denote the unknown velocity $\bm{u}$, pressure $p$, and temperature $T$ fields. The RBVMS formulation of thermal-fluid flows is stated as: $\forall\{\mathbf{w}, q, \eta\} \in \mathcal{W}$, find $\{\mathbf{u}, p, T\} \in \mathcal{V}$, such that
\begin{align}\label{rbvms}
 \nonumber &\int_{\Omega} \mathbf{w}\cdot \rho\left(\mathbf{u}_{,t} + \bm{u}\cdot \nabla \bm{u} - \mathbf{g}\right) d\Omega
 + \int_{\Omega} p \nabla\cdot\mathbf{w} d\Omega \\ \nonumber  
& -\int_\mathbf{\Gamma}\mathbf{w}\cdot\mathbf{h}d\Gamma
 + \int_{\Omega} \nabla^s\mathbf{w} : \mu\nabla^s\bm{u} d\Omega
 + \int_{\Omega} q \nabla\cdot\bm{u} d\Omega \\\nonumber
  &+ \int_{\Omega} \eta \left[  (\rho c_p T)_{,t} + \bm{u}\cdot \nabla (\rho c_pT) + (\rho L f_L)_{,t}  +  \bm{u}\cdot \nabla (\rho L f_L)\right]d\Omega
 +\int_{\Omega}\nabla\eta\cdot\kappa\nabla{T}d\Omega-\int_\Omega\eta Q_T d\Omega\\\nonumber
 & + \Sigma_{e=1}^{n_{el}}\int_{\Omega^e}\tau_M(\bm{u}\cdot\nabla\mathbf{w}+\frac{\nabla {q}}{\rho})\cdot \mathbf{r}_M d\Omega 
 + \Sigma_{e=1}^{n_{el}}\int_{\Omega^e}\rho\tau_C \nabla\cdot\mathbf{w} r_C d\Omega \\  \nonumber 
 &- \Sigma_{e=1}^{n_{el}}\int_{\Omega^e}\tau_M \mathbf{w}\cdot\left[\mathbf{r}_M \cdot \nabla\bm{u}\right]d\Omega-\Sigma_{e=1}^{n_{el}}\int_{\Omega^e}\frac{\nabla\mathbf{w}}{\rho}:(\tau_M\mathbf{r}_M \otimes\tau_M\mathbf{r}_M ) d\Omega\\
 & +\Sigma_{e=1}^{n_{el}}\int_{\Omega^e}\tau_T(\bm{u}\cdot \nabla \eta)  r_T  d\Omega= 0
\end{align}
where $\mathbf{r}_M$, ${r}_C$ and ${r}_T$ are the residuals of momentum, continuity, and energy conservation equations (Eq.~\ref{eq:res}). $\tau_M$, $\tau_C$ and $\tau_T$ are the corresponding stabilization parameters~\cite{yan2018fully}. Other widely used stabilization parameters can be found in~\cite{Tezduyar91c,Takizawa10b, Takizawa12a, Takizawa15a,Takizawa10b, Takizawa12a}. The VMS and its extensions on moving fluid domains using Arbitrary Lagrangian-Eulerian technique (ALE-VMS)~\cite{Bazilevs08a, Takizawa11n} and Space-Time (ST-VMS) technique~\cite{Takizawa11m, Bazilevs13a, Takizawa12e,Takizawa13b, Takizawa14g} have been used as high-fidelity models to simulate a set of challenging fluid dynamics and fluid-structure interaction problems~\cite{takizawa2020a}. Several recent applications include environmental flows~\cite{Korobenko20d}, wind energy~\cite{Bazilevs10a,Takizawa11a,Takizawa11f,Bazilevs13a,Takizawa13a, Takizawa14c, Takizawa14d, Bazilevs14a,Takizawa15b, Korobenko18a}, tidal energy~\cite{Korobenko20c,Korobenko20b, Korobenko20a}, bio-mechanics~\cite{takizawa2014space,terahara2020heart}, and transportation engineering~\cite{kuraishi2019space}.

\begin{figure}[!htbp]
\centering
\includegraphics[scale=0.2]{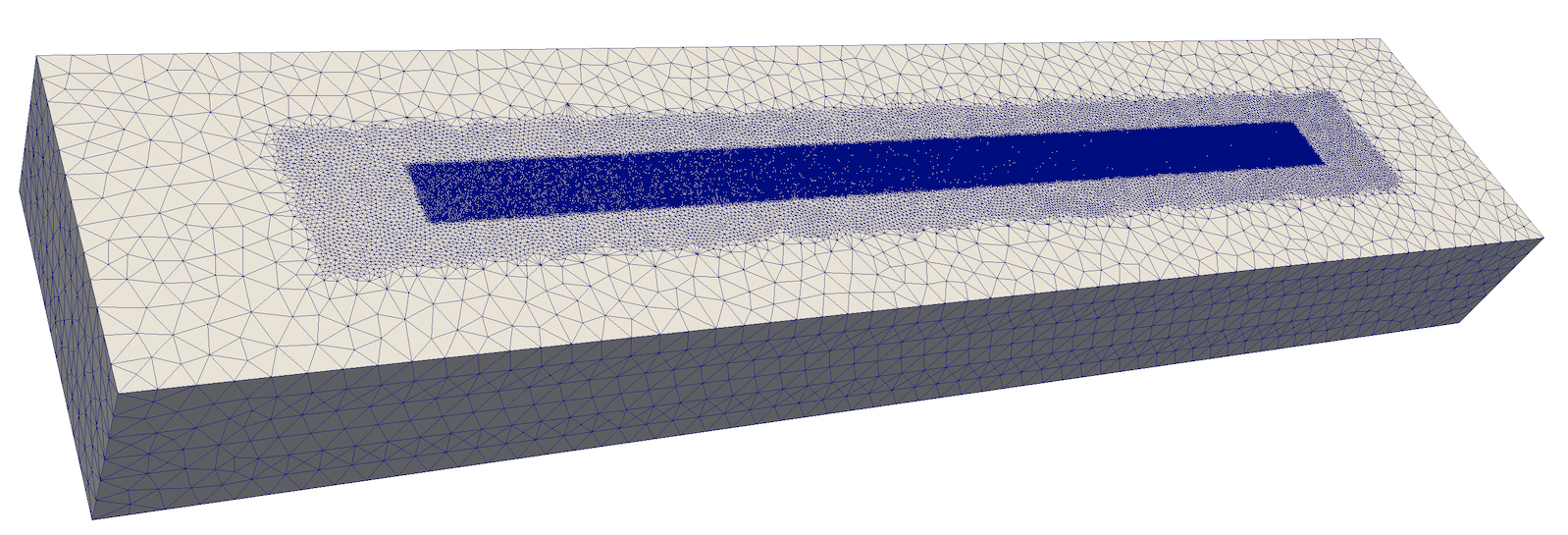}
\caption{The mesh employed in the high-fidelity FEM thermal-fluid simulations.}
\label{fig:FEM_mesh}
\end{figure}
The FEM simulations of the NIST AM-bench test series make use of a mesh with 4,464,276 tetrahedral elements. A local refinement is designed to capture the moving laser, which is shown in Figure~\ref{fig:FEM_mesh}. The unknown velocity, pressure, and temperature are solved in a fully coupled fashion. Generalized-$\alpha$ is used for time integration with $\Delta t = 1\times 10^{-6}$ s. The nonlinear equations are linearized by Newton's method. The resulting linear systems are solved by GMRES with block preconditioning~\cite{saad1986gmres}. These simulations are executed at Stampede2 at Texas Advanced Computing Center (TACC) with 192 CPU processors.

\vspace{0.86cm}

%\noindent \textbf{\large References}

\bibliographystyle{elsarticle-num}
\bibliography{cm_references,cm_references1,references_ROM,main}

\end{document}